\documentclass[report,floatfix,aps,nofootinbib]
{revtex4}
\usepackage{epsfig}

\def\lp {\left( }
\def\rp {\right) }
\def\lb {\left[ }
\def\rb {\right] }
\def\lc {\left\{ }
\def\rc {\right\} }

\def\nn {\nonumber}

\def \kl {\biggl[}
\def \kr {\biggr]}

\def\beq{\begin{equation}}
\def\eeq{\end{equation}}
\def\bea{\begin{eqnarray}}
\def\eea{\end{eqnarray}}
\def\ni{\noindent}

\def\m{\mu}

\def\p{\pi}

\def\s{\sigma}

\begin{document}

\title{Antihyperon Polarization in High-Energy 
Inclusive Reactions}

\author{C.C. Barros Jr. and Y. Hama}

\affiliation{Instituto de F\'{\i}sica, Universidade de S\~{a}o Paulo,\\
C.P. 66318, 05315-970, S\~ao Paulo, SP, Brazil}


\begin{abstract}
We propose a model for the antihyperon polarization in 
high-energy proton-nucleus inclusive reactions, based 
on the final-state interactions between the antihyperons 
and other produced particles (predominantly pions). To 
formulate this idea, we use the previously obtained 
low-energy pion-(anti-)hyperon interaction using 
effective chiral Lagrangians, and a hydrodynamic 
parametrization of the background  matter, which expands 
and decouples at a certain freezeout temperature 
\cite{bjp}. 

\medskip

\ni PACS numbers: 13.75.Gx$\,,$ 13.88.+e 

\end{abstract}

\maketitle

\vspace{5mm}


\section{Introduction}

The unexpected discovery, by Bunce et al. in 1976 
\cite{bunce}, of large $\Lambda$ polarization in the 
inclusive production, induced by proton-Beryllium 
collisions at 200 GeV, opened an interesting field of 
investigation. Earlier experimental data, as those on the 
$\pi p$, $pp$ and $Kp$ collisions \cite{giac}, indicated 
that in general the polarization decreased with the 
energy and, moreover, theoretically it was hard to 
believe that high-energy inclusive reactions could 
produce significant polarization, because it was expected 
that a large cancellation occurred when many uncorrelated 
channels are superposed. 

In the following years many proton-nucleus collision 
experiments have been performed and these experiments 
confirmed the $\Lambda$ polarization, in the $p_t$ 
range of 0-2 GeV \cite{lam,lu}. They also showed that 
in the same domain the ${\overline{\Lambda}}$ 
polarization is very small, consistent with 0. This fact 
created the false idea that only the hyperons could be 
polarized and the antihyperons not. In 1981, Wilkinson 
and his collaborators measured for the first time the 
polarization of a charged hyperon, $\Sigma^+$ \cite{wilk}, 
and found to be positive. In 1983 \cite{deck}, the 
$\Sigma^-$ polarization was measured, with results 
consistent with the $\Sigma^+$ ones, and in 1986, the 
$\Xi^-$ polarization was observed, and it was found to be 
negative \cite{ram}. More recently, polarizations of 
hyperons produced with a $\Sigma^-$ beam also have been 
measured \cite{adam}.

Soon it was possible to obtain polarizations of other 
hyperons: in 1990 the ${\overline{\Xi}^+}$ was measured 
\cite{ho}, negative, and in 1993 Morelos and his 
collaborators measured the ${\overline{\Sigma}^-}$ 
polarization \cite{mor} that is positive. These results 
showed that, refuting the earlier belief, the antihyperon 
polarizations are not always zero. 

Many models have been proposed trying to explain these 
experimental results, such as the Lund model \cite{lund}, 
the quark recombination model \cite{dgr}, and the OPER 
(reggeized one pion exchange) \cite{soff}, that explained 
the $\Lambda$ polarization but were not able to calculate 
the ${\overline{\Lambda}}$ polarization. These models are 
based on the leading-particle effect, that means that the 
observed hyperon is considered as a direct product of the 
incoming proton after a recombination. Clearly, this 
mechanism can be effective only for hyperons but not for 
anti-hyperons. 

Later, models based on constituent quarks \cite{trosh} 
and others \cite{kubo},\cite{anselmino} were proposed, 
but a satisfactory explanation for the antihyperon 
polarization was not achieved. As we can see, the models 
are able to explain the $\Lambda$ polarization (and even 
${\overline{\Lambda}}$), but in fact, all of the models 
predict the production of unpolarized antihyperons and 
this fact shows that these models must be supplemented in 
order to describe also the antihyperon data. 

It is well known that final-state interactions are 
fundamental in the study of the nonleptonic hyperon 
decays, where the final-state strong phases must be 
considered \cite{don}, \cite{kam}, \cite{ta1}. 
In high-energy inclusive processes, a large number of 
particles is produced, so we expect that something 
similar occur in this kind of reactions. In 1993, Y. Hama 
and T. Kodama \cite{hama} used the hydrodynamical model 
to study the effect of final-state interactions of the 
produced (anti) hyperons with the background matter. It 
was supposed that the (anti) hyperons are produced inside 
a medium composed of hot hadronic matter, that is formed 
during the collision, and the polarization is the result 
of the interaction of the hyperon with the surrounding 
matter. In \cite{hama}, this interaction was represented 
by an optical potential and they showed that this 
mechanism could produce polarized hyperons and 
antihyperons. However, some important questions remained 
opened, namely, the origin of the potentials and why 
different potentials are needed for the different (anti) 
hyperons. 

The purpose of this paper is to improve the model 
described in the preceding paragraph, by explicitly 
including some detailed microscopic interaction between the antihyperon and the surrounding thermalized medium
\cite{bjp}. Considering that in this kind of collisions 
pions are predominant among the produced particles, as 
a first approximation, we will include only the 
pion-antihyperon interactions in the study. The observed 
polarization is the average effect of these interactions. 
An important feature is that as the antihyperons and 
pions are in thermal equilibrium inside this fluid, the 
energy involved in the microscopic interactions is 
generally low. Also, since the hyperons are produced 
mostly as the leading particles, we shall confine 
ourselves only to antihyperon polarization. 

A crucial factor to the polarization mechanism is the 
shape of the fluid rapidity distribution, $dN/d\alpha$, 
at the moment of emission. If $dN/d\alpha$ were constant, 
no polarization would be expected because, for 
antihyperons with rapidity $y$, the contributions coming 
from the $\alpha>y$ region would cancel out the ones from 
the $\alpha<y$ domain, as they have opposite signs. But 
in the relevant $\alpha$ domain for the experimentally 
produced antihyperons, $dN/d\alpha$ is decreasing, and 
then, the average polarization is dominated  by the 
contributions from $\alpha<y$. Another remark is that, 
since we are interested in antihyperons which are 
escaping, the microscopic collisions is essentially in 
the forward direction. 

The content of this work is the following. In the next 
section, the previously developed low energy pion-hyperon 
interactions \cite{cy,cm,ccb}, based on chiral 
Lagrangians, are shortly reported. In Sec. III these 
results are used to calculate the polarizations in 
high-energy inclusive processes. In Sec. IV the results 
are shown and, finally, conclusions are drawn in Sec. V. 


\section{Low-Energy Pion-Hyperon Interactions}

In this section, the basic question to be answered is how 
to describe the low energy $\pi \bar{Y}$ interactions in 
a reliable way. The path to be followed is to study the 
$\p Y$ interactions and then to use the $CPT$ invariance 
(neglecting the possible effects of $CP$ violation) to 
determine the amplitudes in the  $\pi \bar{Y}$ scattering. 
We will study the $\p\Lambda$, $\p\Sigma$ and $\p \Xi$ 
interactions, and then, the quantities of interest that 
are cross sections and polarizations, will be calculated. 
The $\bar{Y}$ polarization, for example, is just the $Y$ 
one, with the opposite sign. The cross sections for 
charge conjugate reactions are the same. 

Unfortunately, one must remark that this kind of 
interaction is not very well studied experimentally, the 
first measurement of phase shifts that has been made, was the 
difference $\delta_S-\delta_P$ in the $\p \Lambda$ 
interaction for $\sqrt{s}=m_\Xi$,  by 
the Fermilab E756 experiment \cite{E756}, 
and improved last year by the HyperCP experiment \cite{hyper},
but still with
reasonable experimental uncertainties.  

So, the option is to describe these interactions with a 
model. Some models \cite{nath}-\cite{meis} have been 
proposed to study the $\pi\Lambda$ phase shifts, but 
these values are not accurately determined yet. On the 
other hand, the low energy $\p N$ interactions can be 
described with a high degree of accuracy, with models 
based on  chiral symmetry \cite{leut}, \cite{ellis}. 
So, in this work we will use a model \cite{cy}-\cite{ccb} 
which is based on effective chiral Lagrangians, that 
considers  baryons, spin 3/2 resonances,  rho mesons and 
the $\s$ term. An important feature of the model is the 
inclusion of resonances in the intermediate states. In 
the $\p^+ P$ scattering (3/2 isospin channel) the 
$\Delta^{++}(1232)$ dominates the total cross section at 
low energies and is also very important in the other 
isospin channels. In the $\p Y$ case, we expect that the 
same behavior occurs. The Lagrangians to be considered 
are 
\begin{eqnarray}
&{\cal{L}}_{\pi BB'}& ={g_A\over 2\ f_\pi}\kl B'\gamma_\mu\gamma_5 T^aB \kr
\partial^\mu \phi_a   +   
{\rm h.c.} \label{lbb}            \\
&{\cal{L}}_{\pi BR}& = {g_A^*\over 2\ f_\pi} 
\lc 
{{\overline{R}}^\mu}
\lb g_{\mu\nu} - \lp Z+{1\over 2}\rp \gamma_\mu \gamma _\nu  
\rb
 T^aB \rc
\partial^\nu \phi_a 
  + {\rm h.c.}  \label{lbr}   \\
&{\cal{L}}_{B\rho B'} &= {\gamma_0\over 2}\kl {\overline B'}\gamma_\mu 
T^a B \kr \vec\rho^\mu 
+ {\gamma_0\over 2}\kl {\overline B'} \lp {\mu_{B'}-\mu_{B}
\over 4m_B}\rp i
\sigma_{\mu\nu} T^a\ B \kr .(\partial ^\mu \vec{\rho^\nu} - 
\partial ^\nu \vec{\rho^\mu}) \nn \\
&&  \\
&{\cal{L}}_{\rho\pi\pi}& = \gamma_0\vec\rho_\mu.(\vec\phi\times
\partial^\mu \vec \phi)  
 - {\gamma_0\over 4m_\rho^2}(\partial _\mu 
\vec\rho_\nu - \partial_\nu\vec\rho_\mu).(\partial^\mu \vec\phi\times
\partial^\nu \vec \phi)
   \   \      , 
\end{eqnarray}

\ni where $B$, $R_\m$, $\vec\phi$, $\vec\rho$ are the baryon, the spin 3/2 
resonance, the pion 
and the rho fields with masses $m_B$, $m_r$, $\mu$, and $m_\rho\,$, 
respectively, $\mu_B$  is the baryon magnetic
moment, $T^a$  are the isospin matrices and $Z$ is a 
parameter representing the possibility of the off-shell-resonance having 
spin 1/2.

The first case to be considered is the 
The $\pi\Lambda$ interaction. Since $\Lambda$ has 
isospin 0, the scattering amplitude $T_{\pi\Lambda}$ has the general 
form 
\beq
T_{\pi\Lambda}^{ba}= \overline{u}(\vec{p}\prime ) \lc A + 
{(\not\!k + \not\!k')\over 2}B\rc  \delta_{ba}  
 u(\vec p)      \    \     ,
\eeq

\ni
where $p_{\mu}$ and $p_{\mu}^{\prime}$ are the initial and final 
4-momenta of $\Lambda$, $k_{\mu}$ and 
$k_{\mu}^{\prime}$ are those of the pion.
 Indices $a$ and $b$ indicate the initial and final isospin states 
of the pion.  Fig.~1 shows
the relevant diagrams, where we have omitted 
the crossed diagrams, although included in the calculations.
 To calculate the  diagrams from Fig. 1, we must use
the Lagrangians 
(\ref{lbb}) and (\ref{lbr}), that in this case are

\begin{eqnarray}
{\cal{L}}_{\Lambda\pi\Sigma} &=& {g_{\Lambda\pi\Sigma}\over 2m_\Lambda}\lc
 {\overline\Sigma}
\gamma_\mu \gamma _{5} \vec \tau\Lambda \rc .\partial^\mu \vec \phi  
\ + h.c.
\\
{\cal{L}}_{\Lambda\pi\Sigma^*} &=& g_{\Lambda\pi\Sigma^*} \lc 
{{\overline{\Sigma}^*}^\mu}
\lb g_{\mu\nu} - \lp Z+{1\over 2}\rp \gamma_\mu \gamma _\nu  
\rb \vec\tau\Lambda \rc .\partial ^\nu \vec \phi 
 + h.c.  \  \ , 
  \label{3.1} 
\end{eqnarray}

\begin{figure}[hbtp] 
\centerline{
\epsfxsize=8.cm
\epsffile{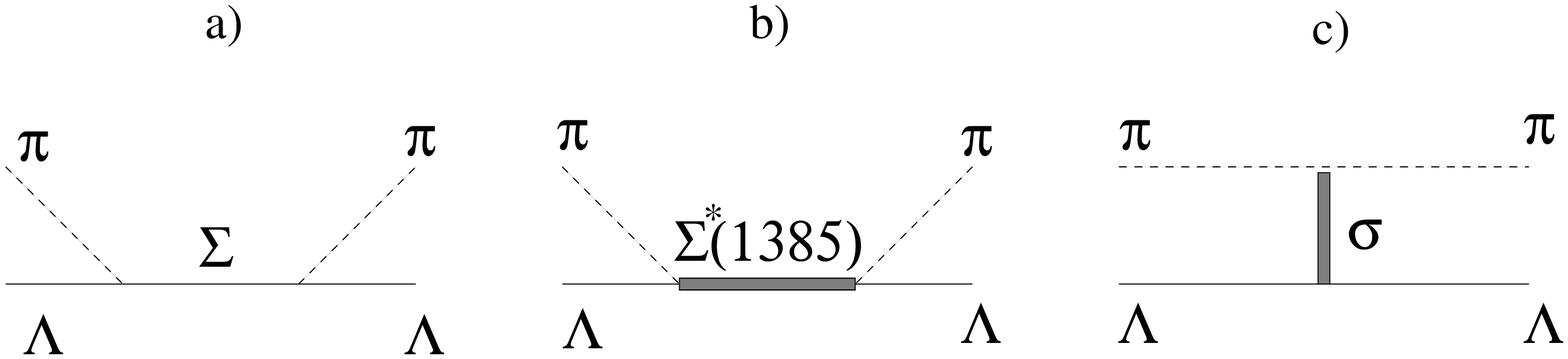}}
\caption{ Diagrams for $\pi\Lambda$ Interaction}
\end{figure}

Calculating
the contribution from Fig. 
1a (intermediate $\Sigma$), we have
\begin{eqnarray}
A_\Sigma &=& {g_{\Lambda\pi\Sigma}^2(m_\Lambda+m_\Sigma)}\lc {\frac{%
s-m_\Lambda^2}{s-m_\Sigma^2}}+{\frac{u-m_\Lambda^2} {u-m_\Sigma^2}}\rc\ ,
 \nonumber \\
B_\Sigma &=& {g_{\Lambda\pi\Sigma}^2}\lc {\frac{m_\Lambda^2-s-2m_%
\Lambda(m_\Lambda+m_\Sigma)}{s-m_\Sigma^2}} 
+{\frac{2m_\Lambda(m_\Lambda+m_\Sigma)+ u-m_\Lambda^2}{u-m_\Sigma^2}}
\rc \ \ .
\end{eqnarray}

\noindent
The interaction with the intermediate
resonance $\Sigma^*$, shown in Fig. 1b, is
\begin{eqnarray}
A_{\Sigma^*} &=& {\frac{g_{\Lambda\pi\Sigma^*}^2}{3m_\Lambda}}\lc 
{\frac{\nu_r}{\nu_r^2-\nu^2}}\hat A-{\frac{m_\Lambda^2+m_\Lambda m_{\Sigma^*}}{m_{\Sigma^*}^2}} 
\lp2 m_{\Sigma^*}^2+m_\Lambda m_{\Sigma^*}
-m_\Lambda^2+2\m^2\rp \right. \nonumber \\
&&
\left.
+{\frac{4m_\Lambda}{m_{\Sigma^*}^2}}\lb (m_\Lambda\!+
m_{\Sigma^*})Z\!+\!(2m_{\Sigma^*}\!+m_\Lambda)Z^2\rb k.k^{\prime}
\rc\ , \nonumber \\
B_{\Sigma^*} &=& {\frac{g_{\Lambda\pi\Sigma^*}^2}{3m_\Lambda}} 
\lc{\frac{\nu}{\nu_r^2-\nu^2}}\hat B - {\frac{8m_\Lambda^2\nu Z^2} 
{m_{\Sigma^*}^2}}\rc\ ,
\end{eqnarray}

\ni
with
\begin{eqnarray}
\hat A&=&{\frac{(m_{\Sigma^*}+m_\Lambda)^2-\m^2}{2m_{\Sigma^*}^2}}
\lc 2m_{\Sigma^*}^3-2m_\Lambda^3-2m_\Lambda m_{\Sigma^*}^2
\right.  \nonumber \\
&&
\left.
-2m_\Lambda^2m_{\Sigma^*}\!\!+\!\m^2(2m_\Lambda\!\!-\!m_{\Sigma^*})
\rc+{\frac{3}{2}}(m_\Lambda\!\!+\!m_{\Sigma^*})t\ ,  \nonumber \\
\hat B &=& {1\over 2m_{\Sigma^*}^2}\lbrack (m_{\Sigma^*}^2-
m_\Lambda^2)^2
-2m_\Lambda m_{\Sigma^*}(m_{\Sigma^*}+m_\Lambda)^2 \nonumber \\
&&
+6\m^2m_\Lambda (m_{\Sigma^*}+m_\Lambda)
-2\mu^2(m_{\Sigma^*}+m_\Lambda)^2+\mu^4\rbrack + {3\over 2}t \ ,  \nonumber \\
&& \label{3.2} 
\end{eqnarray}

\ni
 where  $\nu$ and $\nu_r$ are defined in the 
Appendix A. Note that eq.(\ref{3.2}) shows small differences when compared with
the one shown in
\cite{cy}, where there was a mistake. The correct expression is presented here.

In addition to these contributions,the $\sigma$ term (Fig. 1c) was also
taken into account. In \cite{ols}, \cite{mane}, in the study of $\p N$ 
interactions, it  
 was included just as a parametrization
\begin{eqnarray}
A_{\sigma } &=&a+bt\ , \nonumber \\
B_{\sigma } &=&0\ ,\label{sigma} 
\end{eqnarray}
where $a=1.05\,\mu^{-1}$ and $b=-0.80\,\mu^{-3}$ are constants. 
Recent works \cite{leut}, \cite{man}
consider that the $\sigma$ term may be understood in terms of 
the exchange of two pions, in loop diagrams, 
 as it is shown in Fig 2.

\begin{figure}[hbtp]
\centerline{
\epsfxsize=90.mm
\epsffile{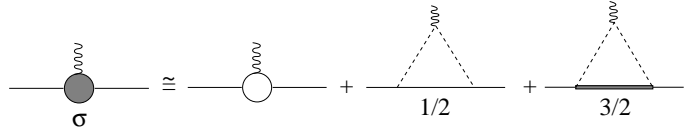}}
\caption{ The scalar form factor  receives contributions from tree
 interactions (white blob)
and triangle diagrams with spin $1/2$ and $3/2$ intermediate states.}
\end{figure}

 So, these ideas lead to a   $\sigma$ contribution of the form
\bea
A_\sigma &=&\sigma_\Sigma+\sigma_{\Sigma^*}      \nn \\
B_\sigma &=&0    \   \  ,
\eea

\ni
where $\sigma_\Sigma$ and $\sigma_{\Sigma^*}$ have been calculated in 
\cite{cm}.

One must remark that  considering the diagrams of Fig. 1, the resulting  
amplitude will be real and consequently the $S$ matrix will not be unitary.
So, the unitarization of the amplitudes must be done. Observing that
the scattering matrix may be expressed as
\beq
M_{ba}={\frac{T_{ba}}{8\pi\sqrt{s}}}=f_{1}+{\frac{(\vec{\sigma}.\vec{k}) 
(\vec{\sigma}.\vec{k}\prime)}{kk\prime}}f_{2} 
\eeq

\ni
the partial wave decomposition can be made with 
\beq
a_{l\pm }={\frac{1}{2}}\int_{-1}^{1}[P_{l}(x)f_{1}(x)+P_{l\pm 1}(x)f_{2}(x)
]\ \ ,
\eeq

\ni
and
the unitarization of  these amplitudes, 
can be made if
we reinterpret them as elements of $K$ matrix and 
write 

\beq
a_{l\pm }^{U}={\frac{a_{l\pm }}{1-ik\ a_{l\pm }}} \ \ . 
\eeq
The phase-shifts are then computed as 
\beq
\delta_{l\pm} = {\rm tg}^{-1}(k\ a_{l\pm}) \ \ ,
\eeq

\ni
and
details on this calculation are shown in \cite{cy}, \cite{ccb}.

The masses and magnetic moments  that we used may be found in \cite{PDG}.
 The $\Lambda\pi\Sigma$
coupling constant is
$g_{\Lambda\pi\Sigma}$=12.92 (see Appendix A). The resonance coupling 
constants were calculated with the comparison of the calculated amplitude
and the Breit-Wigner one, with the data from \cite{PDG} as it was done in 
\cite{cy}. The same procedure was adopted to calculate the other $Y\pi Y^*$
couplings. The obtained value  is $g_{\Lambda\pi\Sigma^*}$ is 9.38
${\rm GeV}^{-1}$.
With these results, we can calculate $\sigma_t$, $d\sigma/d\Omega$ and 
$P$, that will be needed in the next section.

\begin{figure}[hbtp] 
\centerline{
\epsfxsize=8.cm
\epsffile{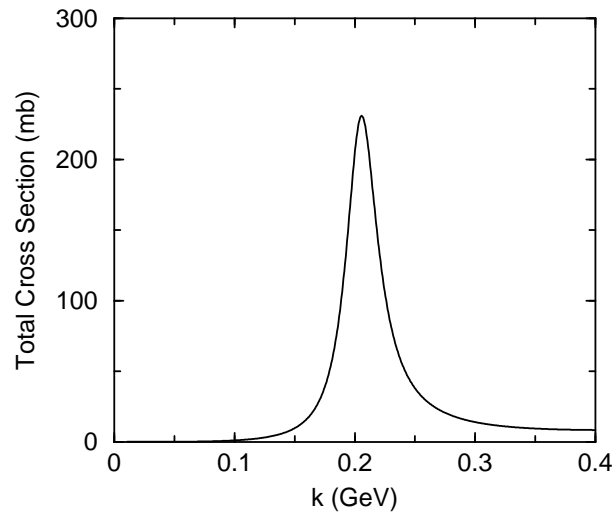}}
\caption{Total $\pi\Lambda$ cross section.}
\label{stl}
\end{figure}

\begin{figure}[hbtp] 
\centerline{
\epsfxsize=8.cm
\epsffile{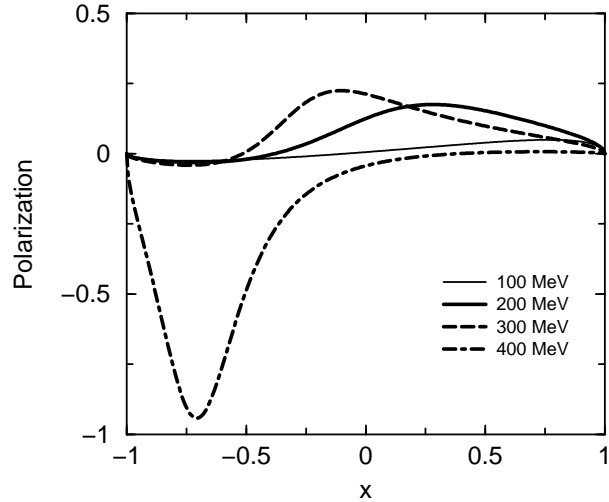}}
\caption{Polarization in the $\pi\Lambda$ interaction, $x=$cos$\theta$.}
\label{pilmpol}
\end{figure}

\ni
Observing  Fig. \ref{stl}, one can see 
a peak in the $\Sigma^*(1385)$ mass, what shows that 
the resonance dominates  the total cross section at low energies.

In the case of $\pi\Sigma$ interaction, the particles have isospin 
1, so the composed system
can have isospin 2,1 or 0. For this reason, the 
scattering amplitude  in this case has the  general form 
\begin{eqnarray}
T_{\alpha\gamma,\beta\delta}&=&
 \langle\pi_\gamma\Sigma_\delta|T|\pi_\alpha\Sigma_\beta\rangle\nonumber \\ 
&=&\overline{u}(\vec{p}\,\prime)\lbrace\lbrack A+  
 {\frac{(\not\!k\,+\!\not\!k^{\prime})}{2}}A^{\prime}\rbrack 
 \delta_{\alpha\beta}\delta_{\gamma\delta} 
+\lbrack B+{\frac{(\not\!k\,+\!\not\!k^{\prime})}{2}}B^{\prime} 
 \rbrack\delta_{\alpha\gamma}\delta_{\beta\delta} 
+\lbrack C+{\frac{(\not\!k\,+\!\not\!k^{\prime})}{2}}C^{\prime} 
 \rbrack\delta_{\alpha\delta}\delta_{\beta\gamma}\rbrace u(\vec p)\ ,
\label{pisig}
\end{eqnarray}

where $\alpha,\,\beta,\,\gamma$ and $\delta$ are isospin indices of the
particles.

\begin{figure}[hbtp]
\centerline{
\epsfxsize=8.cm
\epsffile{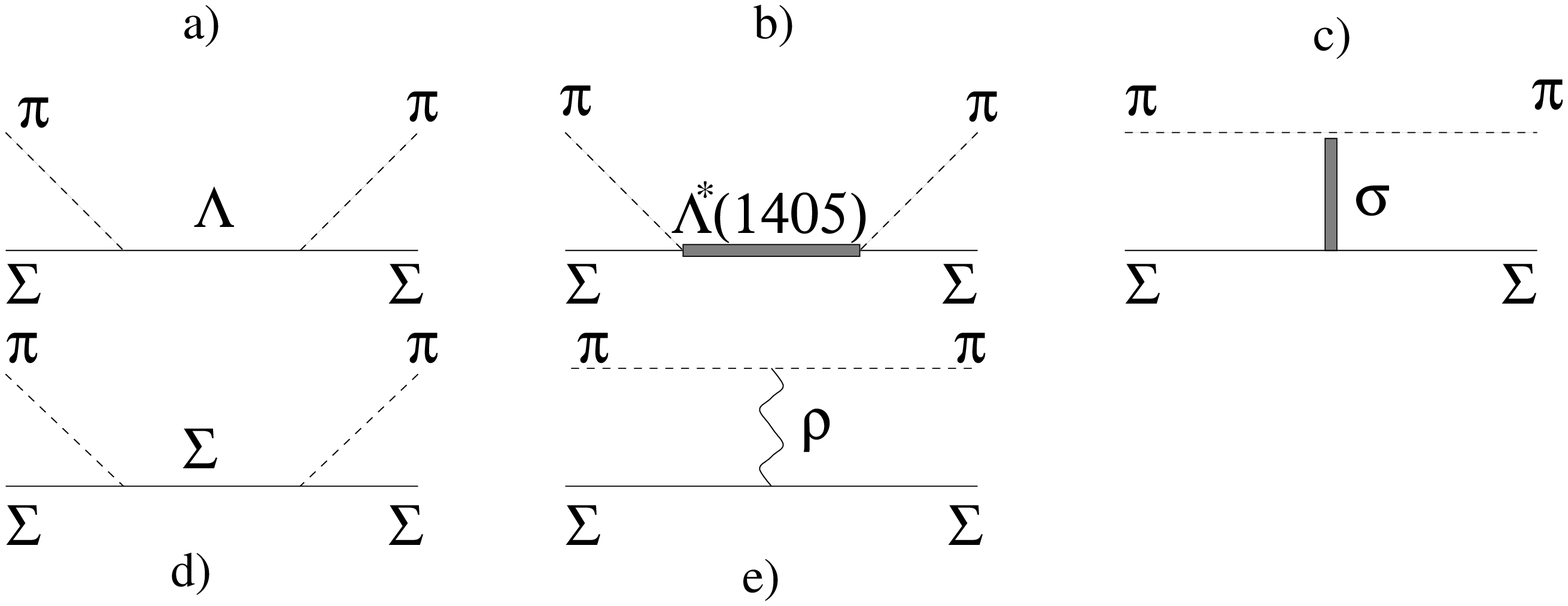}}
\caption{ Diagrams to $\pi\Sigma$ Interaction}
\label{disg}
\end{figure}

The diagrams that we consider for the $\pi\Sigma$ interactions are shown
in Fig. \ref{disg}.
The interactions with the intermediate resonance 
$\Sigma^*(1385)$ could also be included, but
 its decay branching ratio 
 to the $\pi\Sigma$ channel is only 11\%, what may be considered 
as a correction, and then we will neglect it. 

The amplitudes of eq. (\ref{pisig}) for the diagrams of Fig. \ref{disg}
have been calculated in \cite{cy}, and the $\sigma $ contribution is given by
\bea
&& B_\sigma=\sigma_{\Lambda}+\sigma_{\Sigma}+\sigma_{\Lambda^*}  \nn \\
&& A_\sigma=A_\sigma^\prime=B_\sigma^\prime=C_\sigma=B_\sigma^\prime=0 \  \ ,
\eea

\ni
where $\sigma_{\Lambda}$, $\sigma_{\Sigma}$ and $\sigma_{\Lambda^*}$ are
taken from \cite{cm}.

\begin{figure}[hbtp] 
\centerline{
\epsfxsize=8.cm
\epsffile{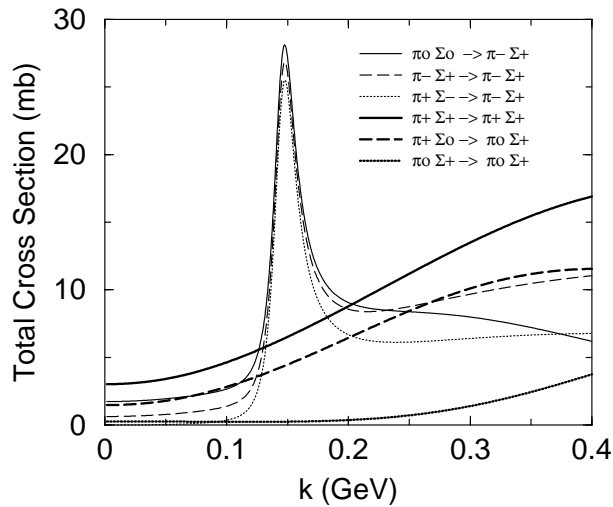}}
\caption{$\pi\Sigma$ cross sections for channels with 
$\Sigma^+$ in the final state.} 
\label{sts}
\end{figure}

\begin{figure}[hbtp]
\epsfxsize=90.mm
\epsffile{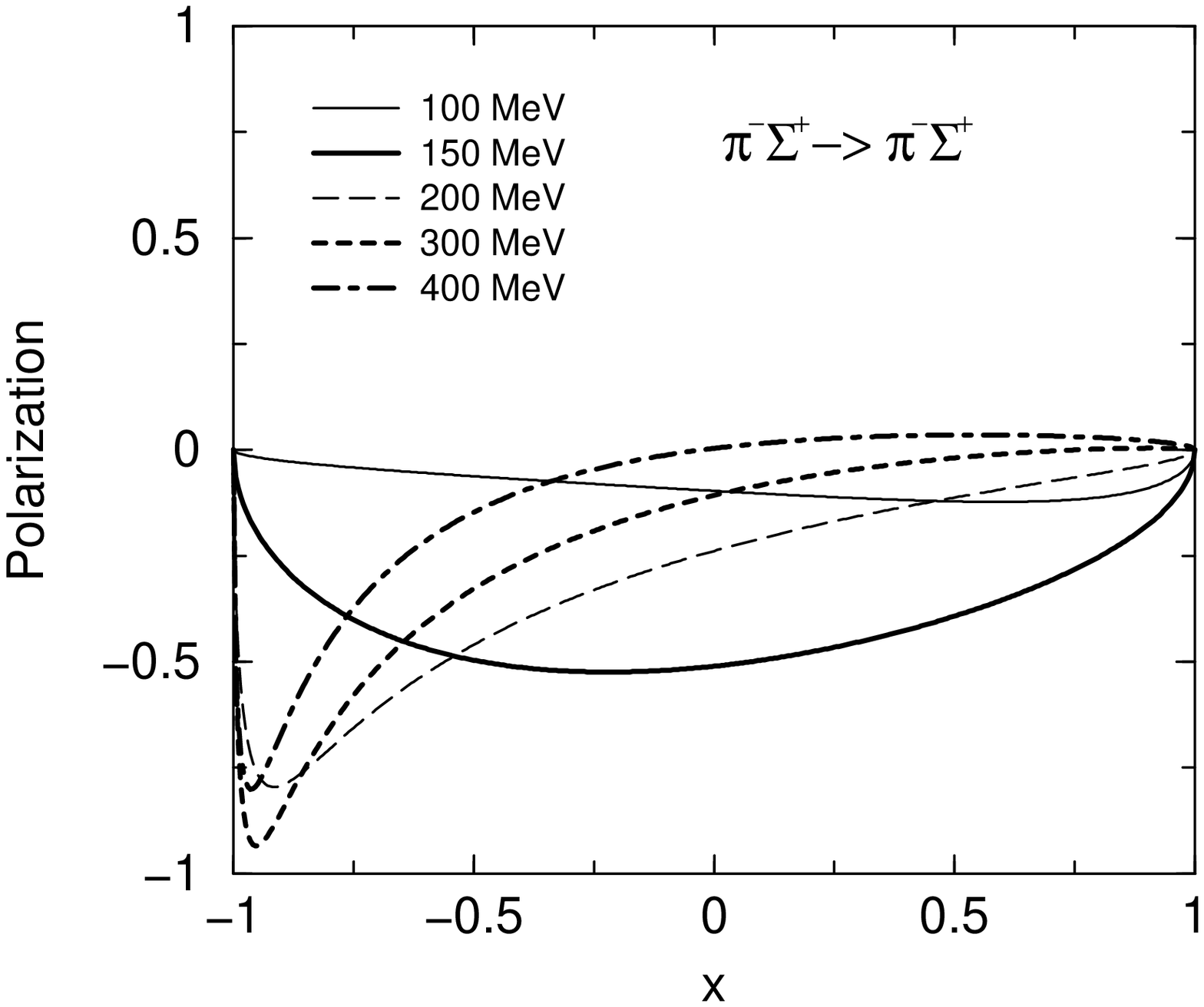}
\epsfxsize=76.mm
\epsffile{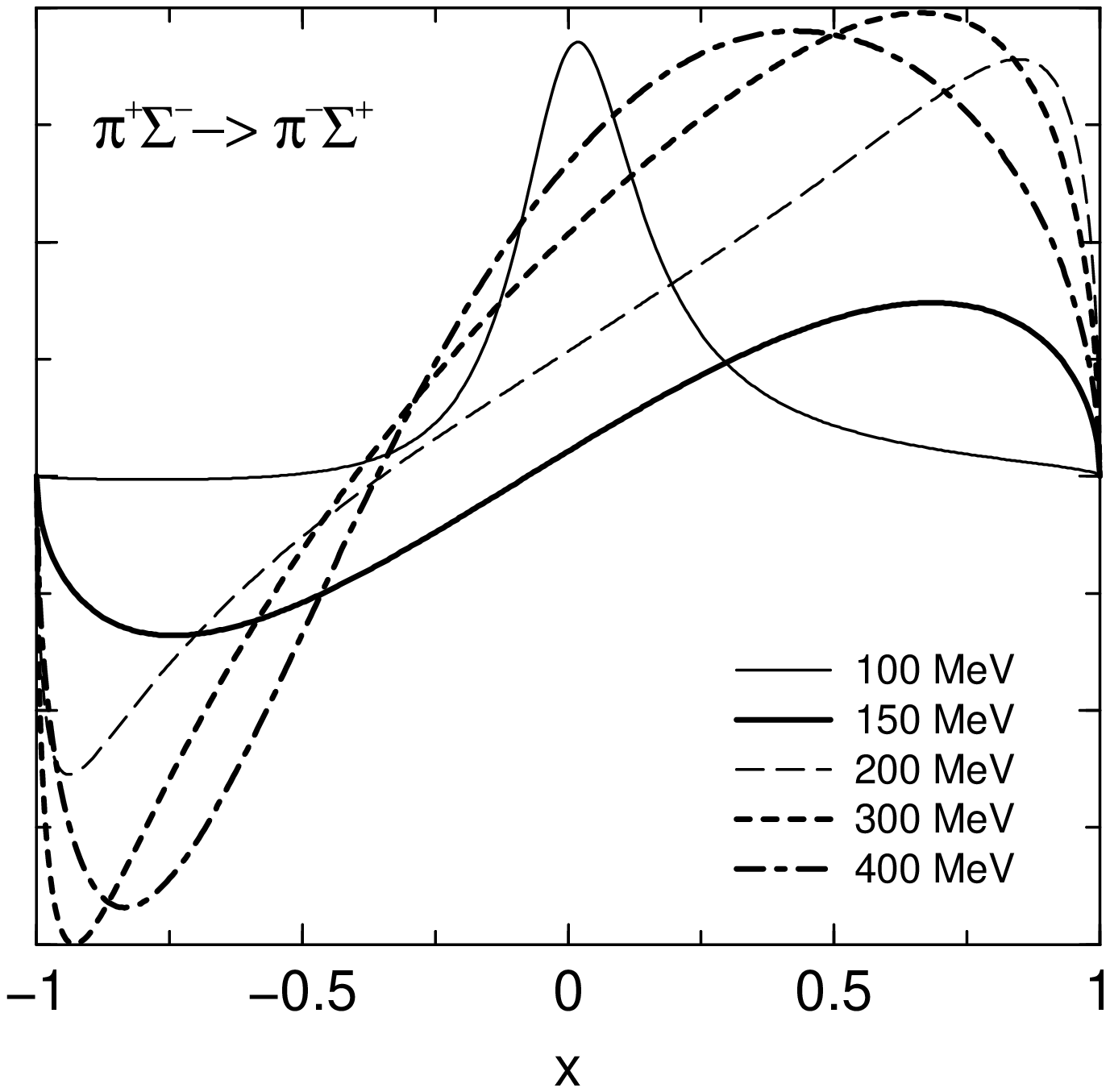}
\epsfxsize=90.mm
\epsffile{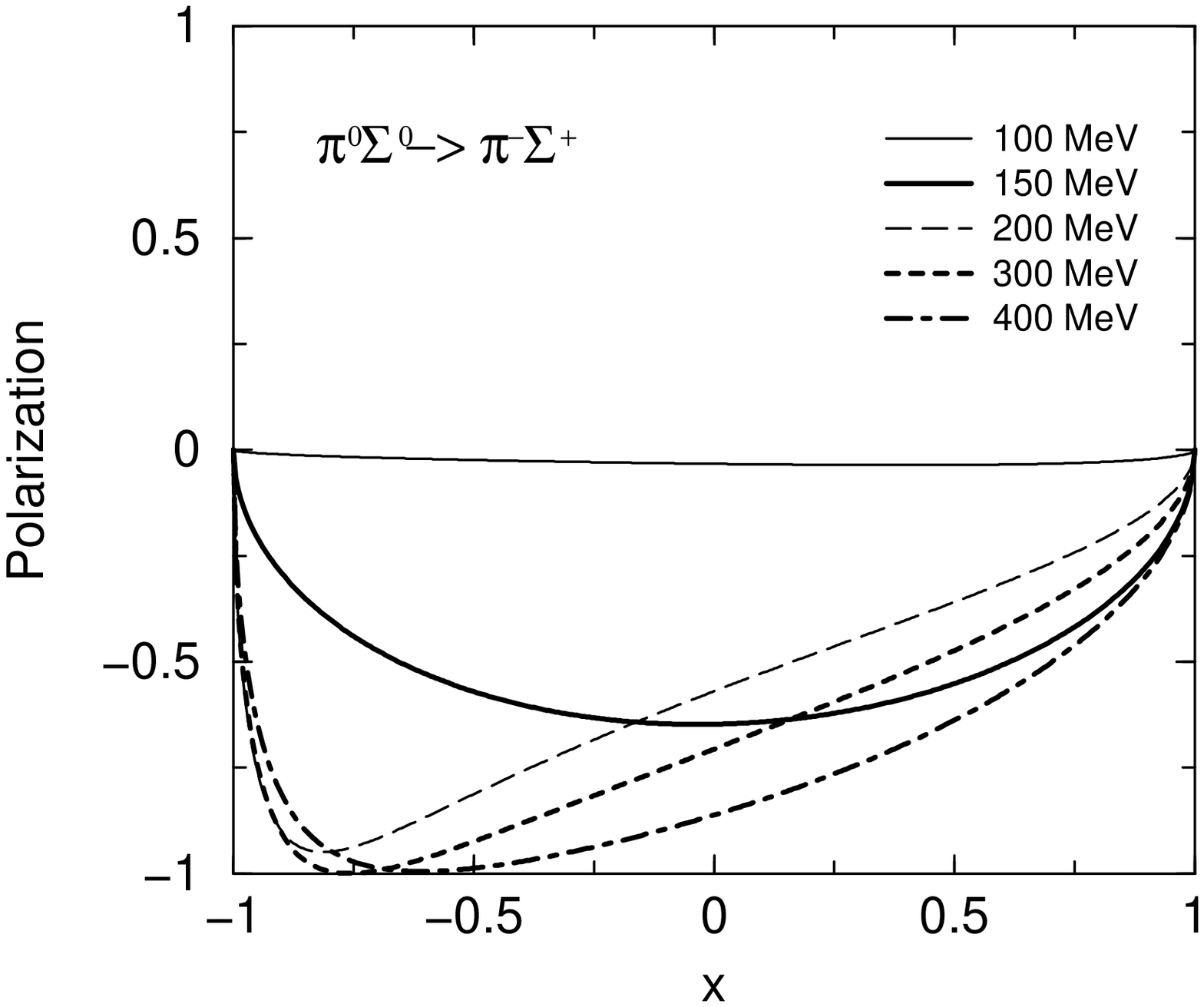}
\epsfxsize=76.mm
\epsffile{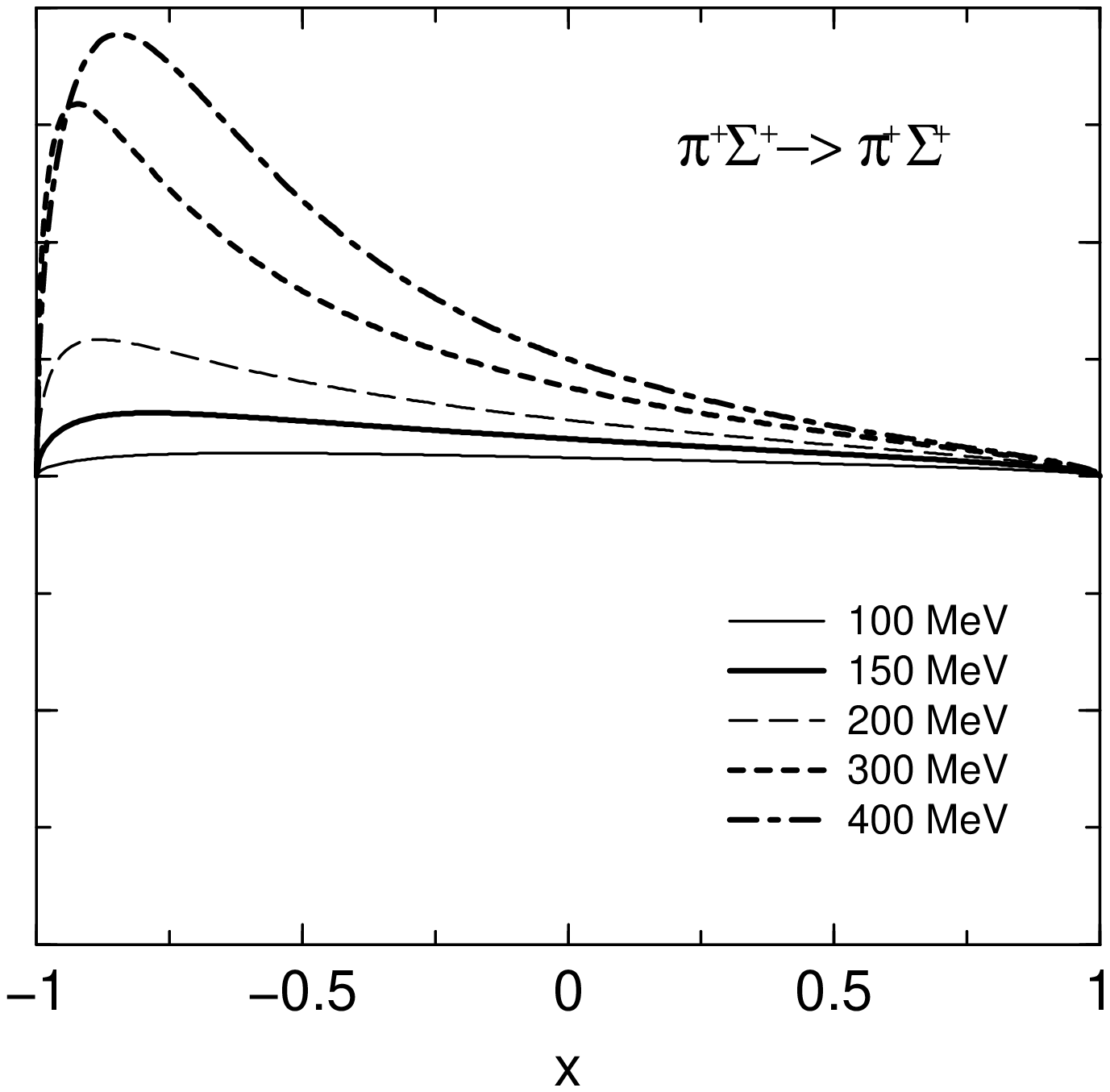}
\epsfxsize=90.mm
\epsffile{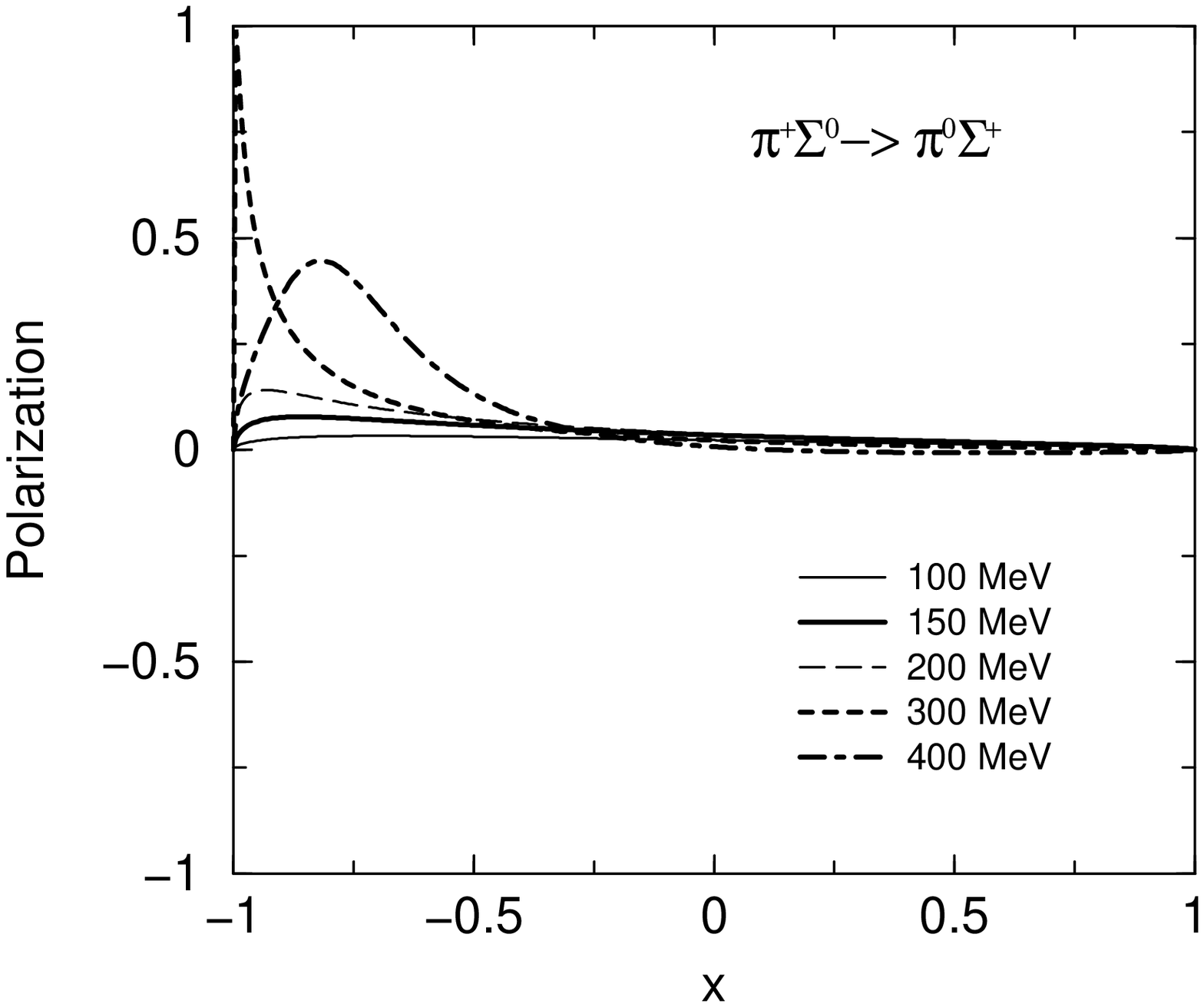}
\epsfxsize=76.mm
\epsffile{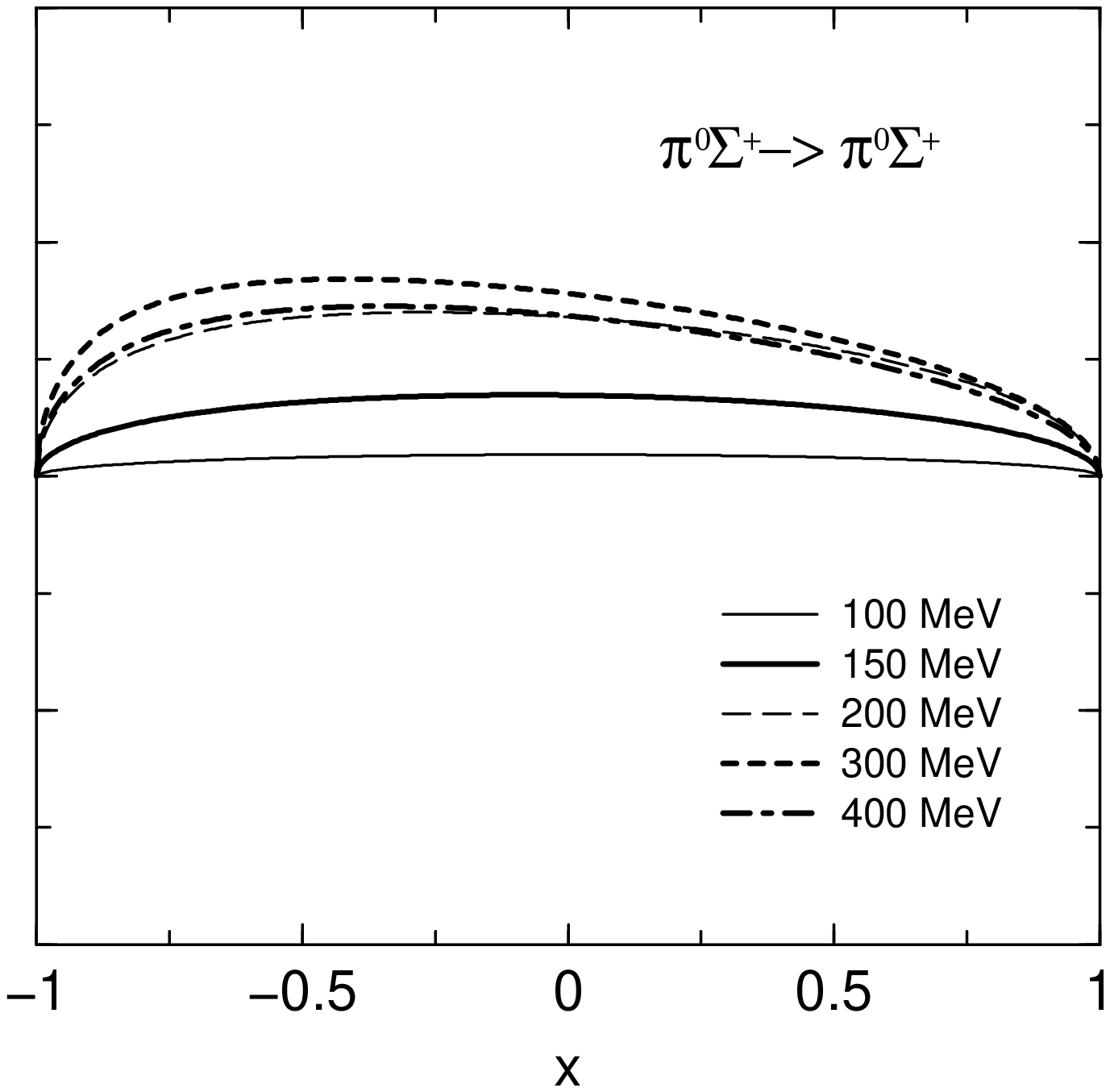}
\caption{Polarizations in the $\pi\Sigma$ Interaction.}
\label{pisgpol}
\end{figure}

The calculated total cross sections with $\Sigma^+$ in the final states
are shown in Fig. \ref{sts}. The resonance, that in this case is 
 $\Lambda^* (1405)$, is still important, but now, the peak is not so high
than it is in the other cases  (less than 30 mb). It appears in the reactions
where the 0 isospin state is important. The polarizationsfor these reactions 
are shown in Fig. \ref{pisgpol}.

Now, let us turn our attention to the $\p \Xi$ interaction.
This case is very similar to the $\pi N$ scattering, because $\Xi$ 
has isospin 1/2 (as the nucleon) and the main difference is that the 
resonance of interest $\Xi^*(1533)$ has isospin $I$=1/2 (instead of $I$=3/2 
as $\Delta(1232)$). Then, the scattering amplitude $T_{\pi\Xi}^{ba}$ has the 
general form 
\begin{eqnarray}
T_{\pi\Xi}^{ba}&=&\overline{u}(\vec{p}\,\prime)\lbrace\lbrack A^+ +
 {\frac{(\not\!k+\!\!\not\!k^{\prime})}{2}}B^+\rbrack\delta_{ba}\nonumber \\
 &&+\lbrack A^- +{\frac{(\not\!k+\!\!\not\!k^{\prime})}{2}}B^-\rbrack
i\epsilon_{bac}\tau^c\rbrace u(\vec p)\ .
\end{eqnarray}

\begin{figure}[hbtp]
\centerline{
\epsfxsize=8.cm
\epsffile{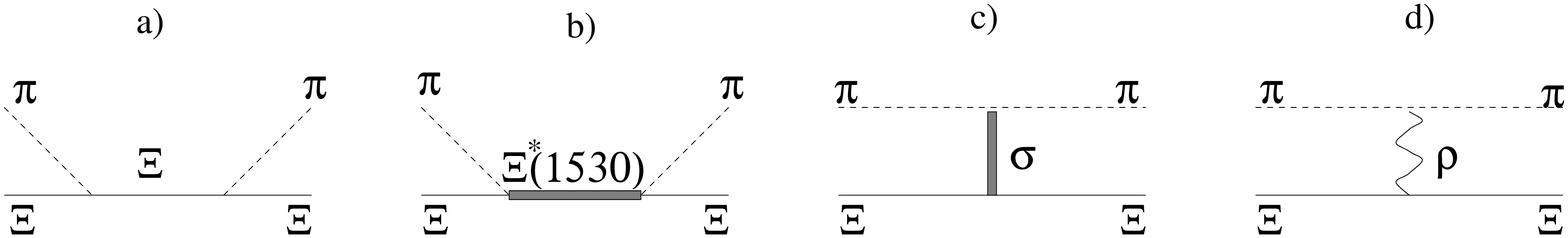}}
\caption{Diagrams to $\pi\Xi$ Interaction}
\label{pix}
\end{figure}

The calculated $A^\pm$ and $B^\pm$ for the diagrams of Fig. \ref{pix} may
be found in \cite{cy},\cite{ccb}.

\begin{figure}[hbtp] 
\centerline{
\epsfxsize=8.cm
\epsffile{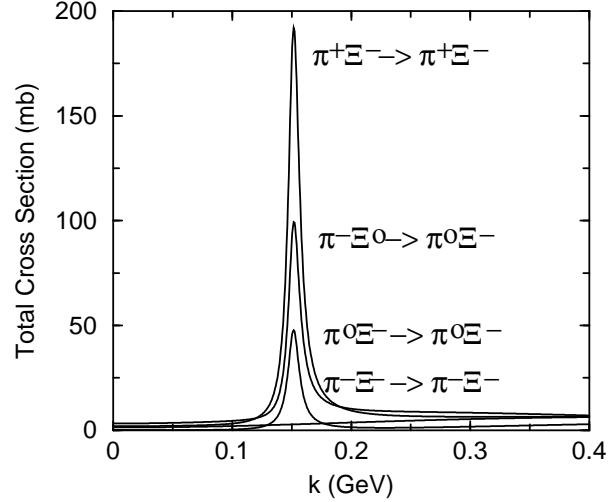}}
\caption{$\pi\Xi$ cross sections for channels with $\Xi^-$ 
in the final state.} 
\label{xst}
\end{figure}

\begin{figure}[hbtp]
\epsfxsize=90.mm
\epsffile{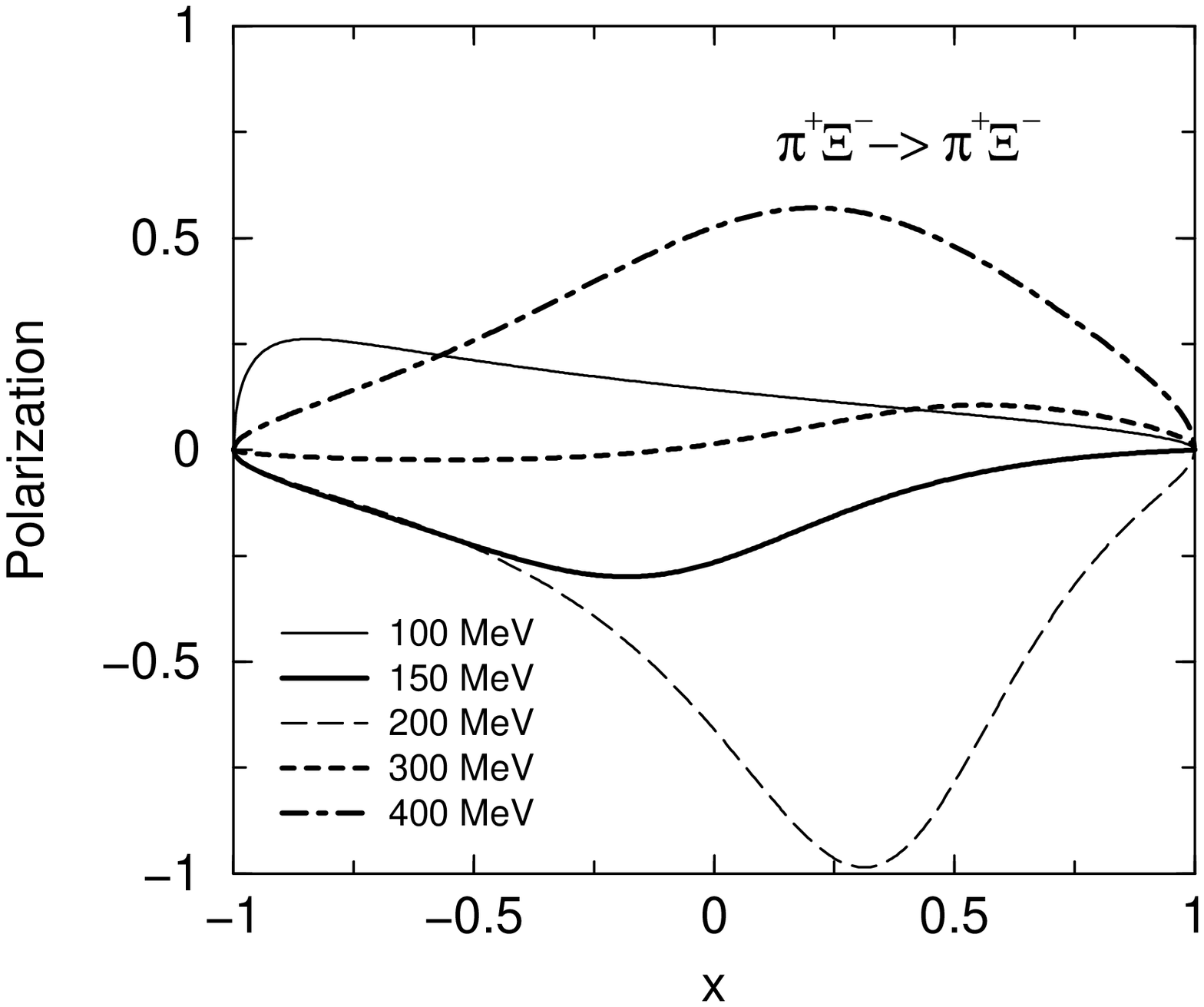}
\epsfxsize=74.mm
\epsffile{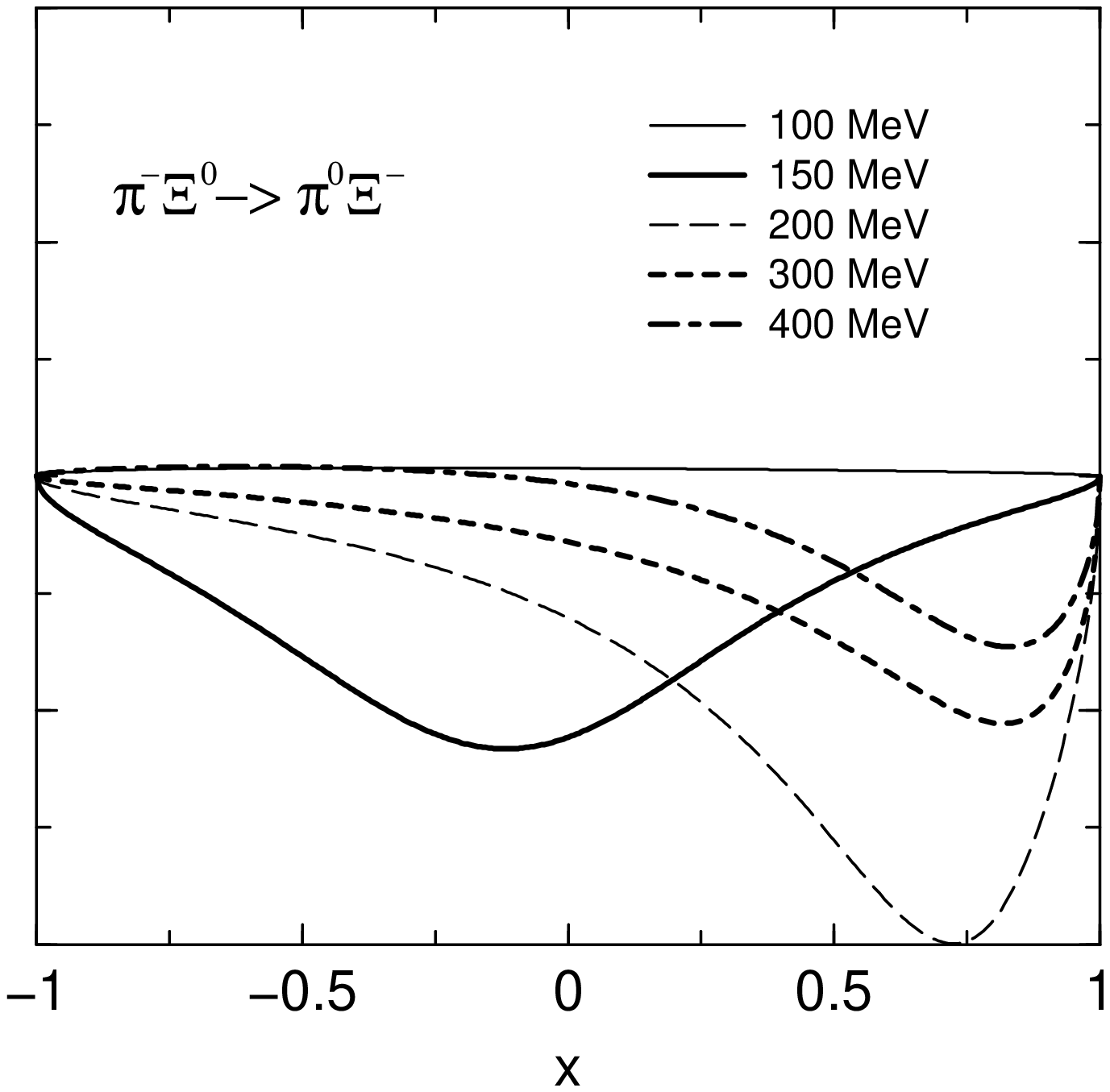}
\epsfxsize=94.mm
\epsffile{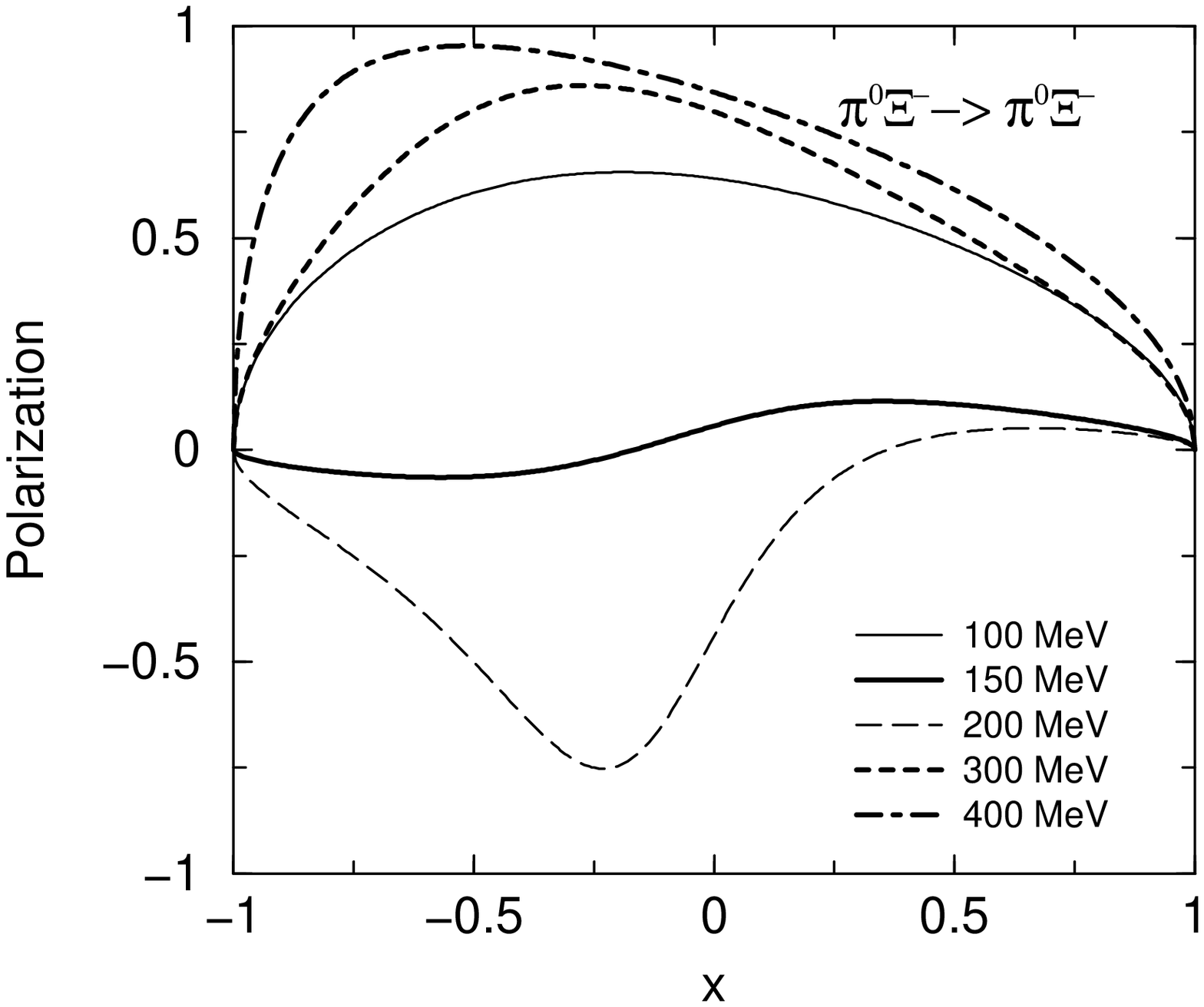}
\epsfxsize=76.mm
\epsffile{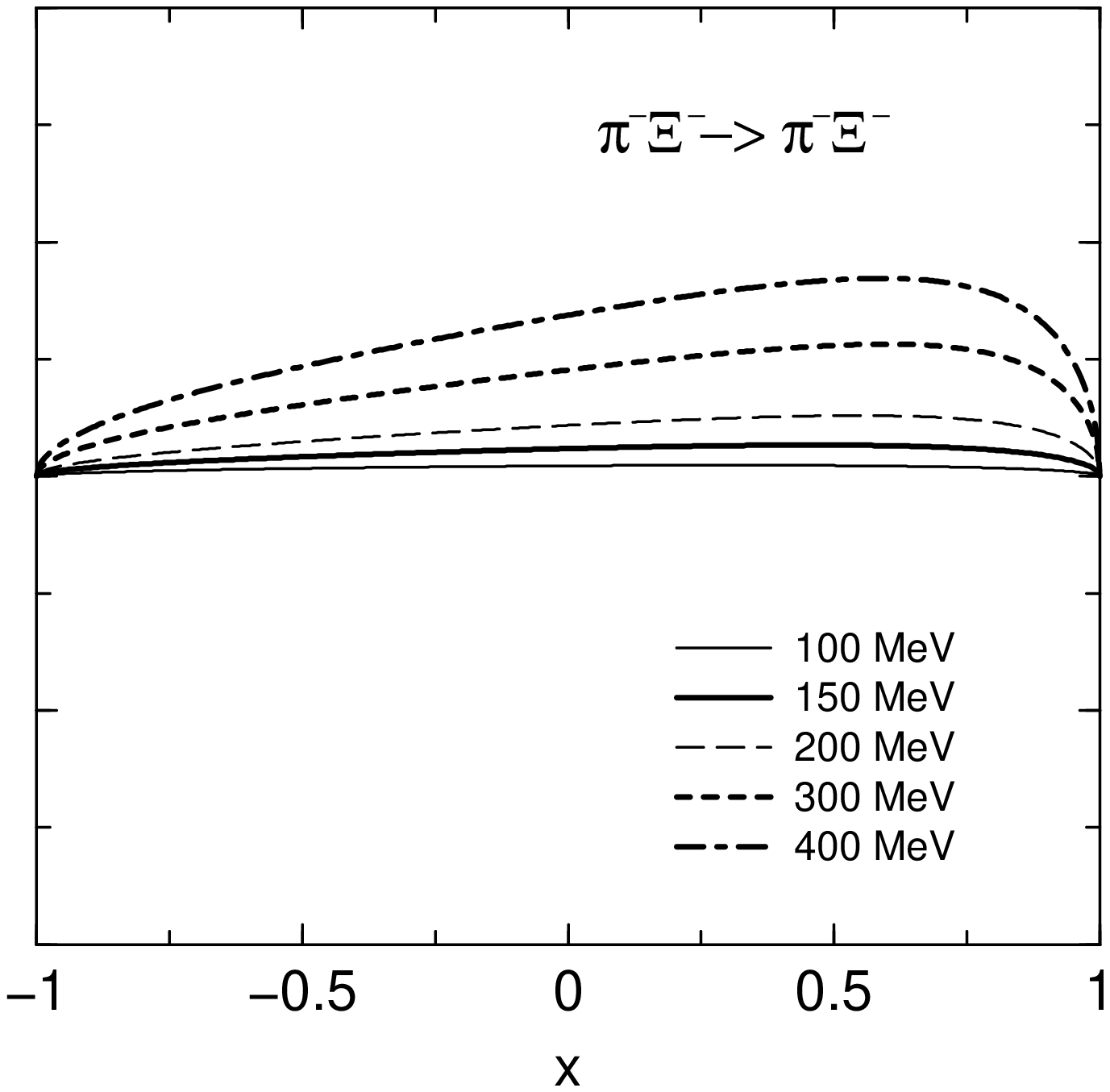}
\caption{Polarizations in the $\pi\Xi$ Interaction.}
\label{pixpol}
\end{figure}

In Fig. \ref{xst}, we can see that one more time the resonance is very
important, now the $\Xi(1533)$ resonance contribution
dominates three of the reactions.

So, now we have the $\p \Lambda$, $\p\Sigma$ and $\p \Xi$ scattering 
amplitudes, that are the basic elements to calculate the polarization in 
high-energy processes.


\section{Polarization in Inclusive Processes}

In this section, it will be shown how to calculate the $\bar{Y}$ polarization
in high energy processes. As it was said, we consider that during the 
collision, a system composed of expanding hot hadronic matter is formed.
This system may be considered to be a fluid, composed of elements, that expands
according to a Gaussian law \cite{yf},\cite{yxt}
\beq
{1\over {\rm ch}\alpha_t\ {\rm sh}\alpha_t}{d^2\sigma\over d\alpha \
  d\alpha_t d\phi}= A\lb e^{-\beta (\alpha -\alpha_0)^2}+
e^{-\beta (\alpha +\alpha_0)^2}     \rb e^{-\beta_t\alpha_t^2}  \  \  ,
\label{dnda}
\eeq

\ni
where $\alpha$ and $\alpha_t$ are the longitudinal and transverse rapidities 
of these elements of fluid, $\phi$ is the azimutal angle and $\beta$, 
$\beta_t$ and $\alpha_0$ are 
parameters that determine the shape of this distribution.
The values of these parameters depend on the participants of the collision
(proton, beryllium, ...) and on their energies.

The pions and hyperons are produced thermally inside these fluid elements, 
with initial momenta (relative to the fluid element)
 $\vec\pi_0'$ (pions) and $\vec\Lambda_0'$ (hyperons) and energies
 $E_{\pi_0}'$, $E_0'$,
obeying the statistical distributions
\beq
F_{BE}(\pi_0')\propto{1\over \exp({E_{\pi_0}'\over T}) -1}  \  \   ,
\eeq

\beq
F_{FD}(\Lambda_0')\propto{1\over \exp({E_0'\over T}) +1} \  \   .  
\eeq

\ni
The hyperon emerges with final momentum
$\vec\Lambda'$,  energy $E'$ and polarization ${\vec P}^\prime$.

The produced hyperons interact, and as 
the relative energy between the $\p$ and the
$Y$ is small, the amplitudes
obtained in the last section may be used to calculate
  the polarization $\vec P^\prime$
and $d\sigma/dt$.  
Then the average polarization may be calculated by the expression 
\beq
\langle \vec P \rangle = {\int \left\{ \left(\vec P^\prime {d\sigma\over dt}
\right)_{R_1} + ...+\left(\vec P^\prime 
{d\sigma\over dt}\right)_{R_N}
\right\} {\cal{G}} d\tau \over 
\int \left\{ \left( {d\sigma\over dt}
\right)_{R_1} + ...+\left( 
{d\sigma\over dt}\right)_{R_N}
 \right\} {\cal{G}}d\tau}         \   \   ,
\label{pm}
\eeq

\noindent
that considers the  elastic and with charge exchange
 $R_i$ reactions,
in the interactions of the hyperons with 
 $\pi^+$, $\pi^-$ and $\pi^0$.

The factor ${\cal{G}}$ that appears in eq. (\ref{pm}) contains the 
statistical weighs of the production of the particles and the ones relative to 
the expansion of the fluid, and
can be written as
\bea
{\cal{G}}=&&{(d^2N /d\alpha \ d\alpha_t) \over 
\left(\exp\left({E_{\pi_0}'\over T}\right)-1\right)
\left(\exp\left({E_0'\over T}\right)+1\right)} 
\Lambda _0'^2\pi _0'^2 \nn \\
&&\times
\delta \left(E_0'+E_{\pi _0}'-E'-\sqrt {m_\pi ^2 + (\vec \pi_0'+
\vec \Lambda _0'-\vec \Lambda ')^2}\right)  
\  \   , \nn \\
\eea
\noindent
and the eight dimension integration element is
\beq
d\tau= d\alpha\ d\alpha_t  \ d\vec \Lambda _0'd\vec\pi _0'  \  \  .
\eeq

In order to calculate the average polarization (\ref{pm}),
we must know the parameters $\beta$, $\beta_t$ and $\alpha_0$ 
of the rapidity distribution (\ref{dnda})  of the 
 fluid elements and if this expression is a good choice.
 It is possible to answer these questions if we use the experimental data of 
the rapidity distributions of the pions produced in proton-proton collisions 
\cite{bell} (that is a collision similar to the $p-$Be).
In the hydrodynamical framework, the longitudinal rapidity distribution of 
the produced
pions is given by
\beq
{d\sigma\over dy}=\int {d\sigma\over d\alpha}(\alpha) \  
{d\sigma\over dy'}(y-\alpha)  \  d\alpha  \  \  ,
\label{conv}
\eeq

\ni
that is
 a convolution of the rest-frame
rapidity distribution $d\sigma/d y'$,
 and the fluid rapidity distribution, 
 $d\sigma/d\alpha$.

The pion
momentum distribution with respect to the fluid element 
(considering that they are produced thermally) is
\beq
{d\sigma \over d\vec p_\pi^\prime}= 
{1\over E_\pi '}{d\sigma\over dy'd\vec p_t^\prime}=
{1\over e^{E_\pi '\over T}-1}
\eeq

\ni
that after some manipulations  gives
\beq
{d\sigma\over dy'}\sim Ce^{-\beta'y'}
\eeq

\ni
with $\beta'\sim 0.98$ for the temperature $T\sim m_\pi$.

The distribution
 $d\sigma/d\alpha$ may be obtained from eq. (\ref{dnda}), and will be a 
sum of Gaussian functions
\beq
{d\sigma\over d\alpha}= A^\prime \lbrack 
e^{-\beta(\alpha-\alpha_0)^2}+e^{-\beta(\alpha+\alpha_0)^2}   \rbrack 
 \  \  ,
\eeq  

\begin{figure}[hbtp]
\epsfxsize=8.cm
\epsffile{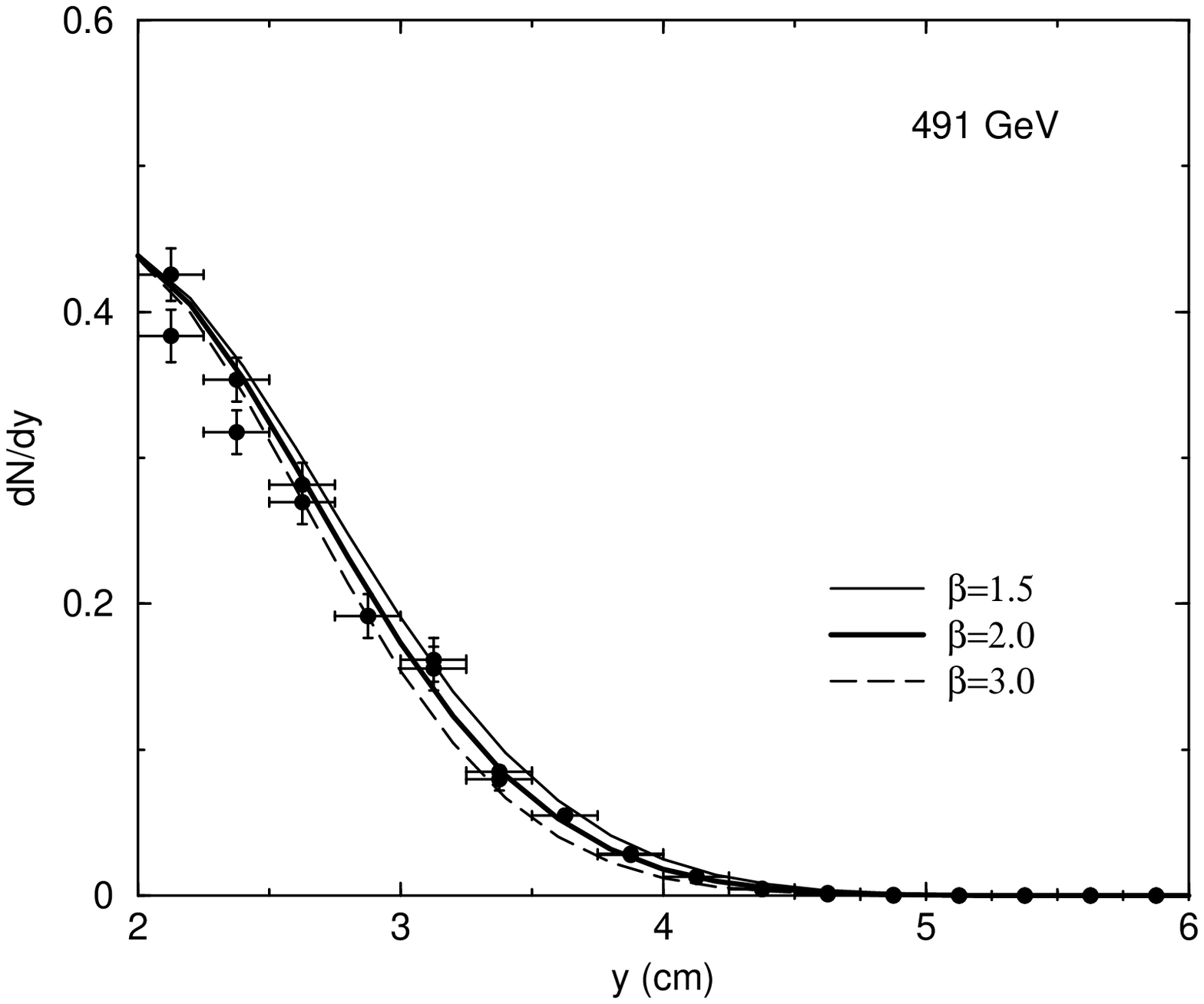}
\epsfxsize=8.cm
\epsffile{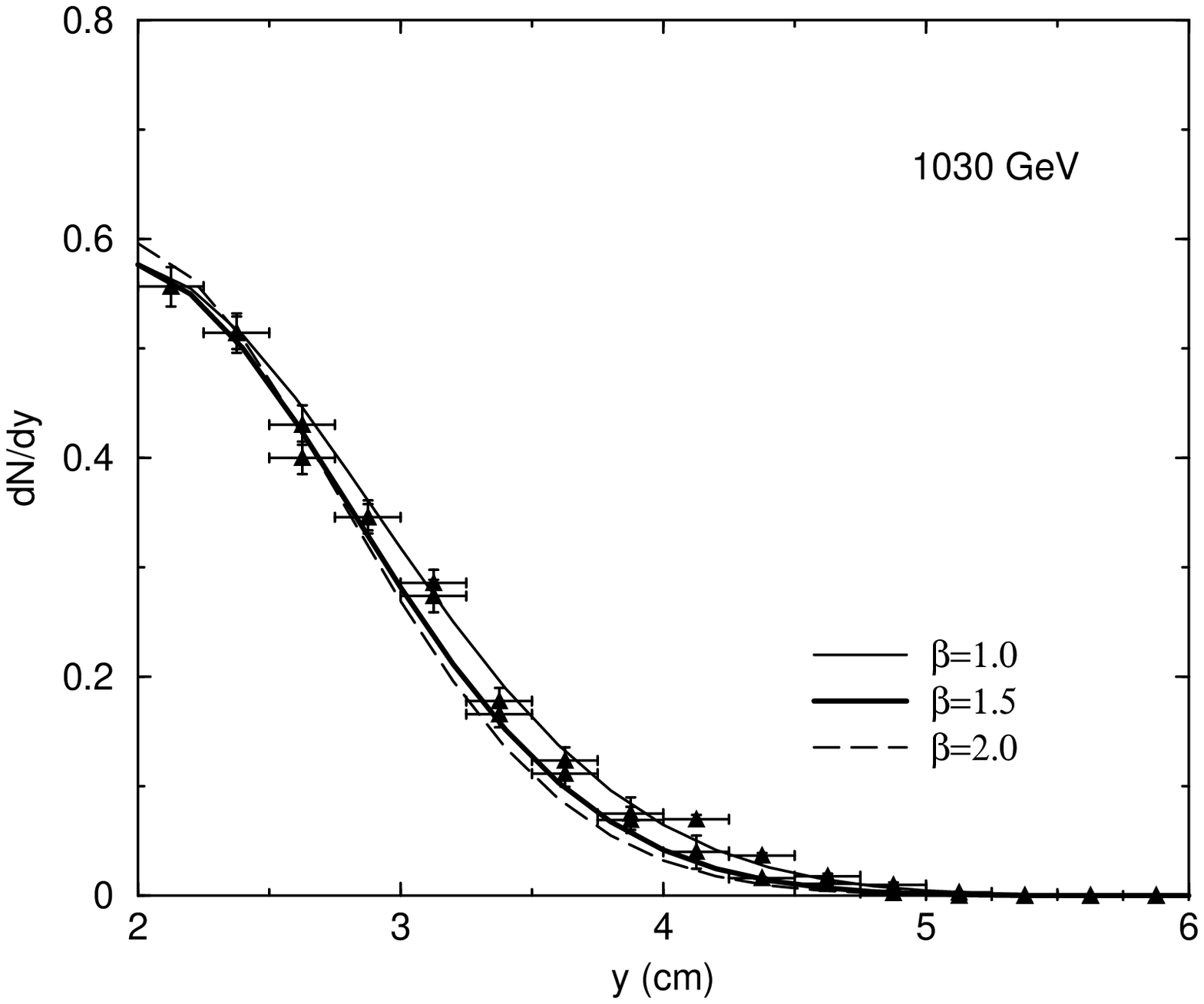}
\caption{Parametrization of
$dN/d y$  for $pp$ collisions compared
with the data from \cite{bell}.}
\label{isrd}
\end{figure}

\ni
 inserting this expression in  (\ref{conv}), integrating and 
fitting
the parameters
$A^\prime $, $\beta$ and $\alpha_0$, we can compare the results with the 
experimental data from ISR 
  \cite{bell} . The results are shown in Fig. \ref{isrd}.

\begin{figure}[hbtp]
\epsfxsize=8.cm
\epsffile{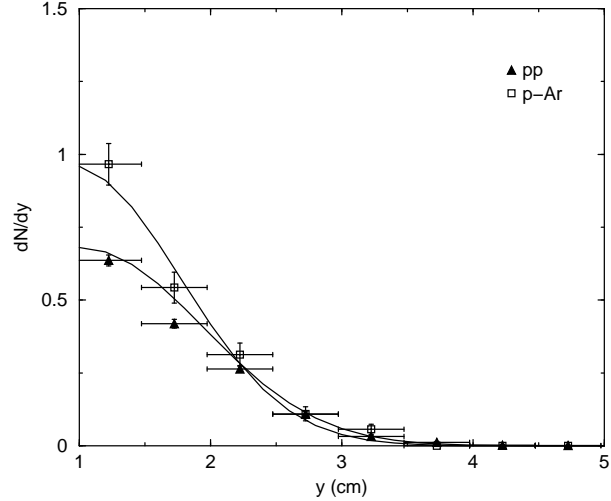}
\caption{$d N/d y$ for 200 GeV $pp$ and $pAr$ 
collisions \cite{marz}.}
\label{dnpa}
\end{figure}

The best fits for $d\sigma/ dy$ are  obtained with the parameters 
$A^\prime=0.45$, $\alpha_0=1.8$ and $\beta=2.0$
for 491 GeV ($\sqrt{s_{NN}}$=31 GeV) and 
$A^\prime =0.58$, $\alpha_0=1.9$ and $\beta=1.5$
for 1030 GeV ($\sqrt{s_{NN}}$=44 GeV) incident protons.
Our real interest is for 330, 400 and
   800 GeV incident protons, that are the values for which there are
antihyperon polarization data.
The respective parameters may be obtained by
interpolation. 
 Observing Fig. \ref{isrd} one can see that the 
$\beta$ parameters may vary from 1.0 to 3.0
 without changing significantly the accord with the experimental data.
So  in 
the calculation of the
average polarization, we will consider the parameter $\beta$ inside this range.
One must remark that for heavier targets, the value of the parameters may 
change, as it can be seen in Fig. \ref{dnpa}, where  $\beta$=4.0 is reasonable
for $p-Ar$ collisions ($m_{Ar}$=39.95) and 2.0 for $pp$ collisions.

The transverse  rapidity parameter $\beta_t$ will be assumed to be 
\cite{yf}
 $\beta_t=6.6$ for 800 Gev,  7.5 for 400 GeV and 8.0 for 330 GeV.

The reactions $R_i$ to be considered to calculate the ${\overline{\Lambda}}$
polarization with eq. (\ref{pm}) are
\begin{eqnarray}
\pi^+{\overline{\Lambda}} &\rightarrow &\pi^+{\overline{\Lambda}} \nonumber \\
\pi^-{\overline{\Lambda}} &\rightarrow &\pi^-{\overline{\Lambda}} \nonumber \\
\pi^0{\overline{\Lambda}} &\rightarrow &\pi^0{\overline{\Lambda}}  \    \   .
\end{eqnarray}

\ni
To calculate the ${\overline{\Sigma}}^-$ polarization,
we consider six reactions (elastic and charge exchange), that are
the charge conjugate  of the ones shown in Fig. \ref{sts}
\begin{eqnarray}
\pi^-{\overline{\Sigma}}^-&\rightarrow &\pi^-{\overline{\Sigma}}^-
 \nonumber \\
\pi^0{\overline{\Sigma}}^-&\rightarrow &\pi^0{\overline{\Sigma}}^- \nonumber \\
\pi^+{\overline{\Sigma}}^-&\rightarrow &\pi^+{\overline{\Sigma}}^- \nonumber \\
\pi^-{\overline{\Sigma}}^-&\rightarrow &\pi^+{\overline{\Sigma}}^- \nonumber \\
\pi^-{\overline{\Sigma}}^-&\rightarrow &\pi^0{\overline{\Sigma}}^- \nonumber \\
\pi^0{\overline{\Sigma}}^-&\rightarrow &\pi^+{\overline{\Sigma}}^-
\end{eqnarray}

\ni
It is important to remark that even at low energies, the 
 exchange reactions $\pi\Lambda\rightleftharpoons\pi\Sigma$ may occur. 
However, observing the fact that the resonance diagrams dominate the cross 
sections in this region, and the facts that the  
$\Lambda^*$  resonances   may not be formed
in $\pi \Lambda$ interactions, and that
$g_{\Sigma\pi\Sigma^*} <<  g_{\Lambda\pi\Sigma^*}$ 
(considering $\Sigma^* $(1385)) we expect that the 
amplitude of the $\pi\Lambda \rightarrow \pi\Sigma$ reactions are not larger 
then 20\% \cite{cy}  of the
ones for the  $\pi\Sigma \rightarrow \pi\Sigma$, so, in a first approximation 
this kind of reactions may be neglected.

In the ${\overline{\Xi}}^+$ case, four reactions are considered,
\begin{eqnarray}
\pi^-{\overline{\Xi}}^+&\rightarrow &\pi^-{\overline{\Xi}}^+ \nonumber \\
\pi^0{\overline{\Xi}}^+&\rightarrow &\pi^0{\overline{\Xi}}^+ \nonumber \\
\pi^+{\overline{\Xi}}^+&\rightarrow &\pi^+{\overline{\Xi}}^+ \nonumber \\
\pi^+{\overline{\Xi}}^+&\rightarrow &\pi^- {\overline{\Xi}}^+ \  \  .
\end{eqnarray}


\section{Results}

\begin{figure}[hbtp]
\centerline{
\epsfxsize=8 cm
\epsffile{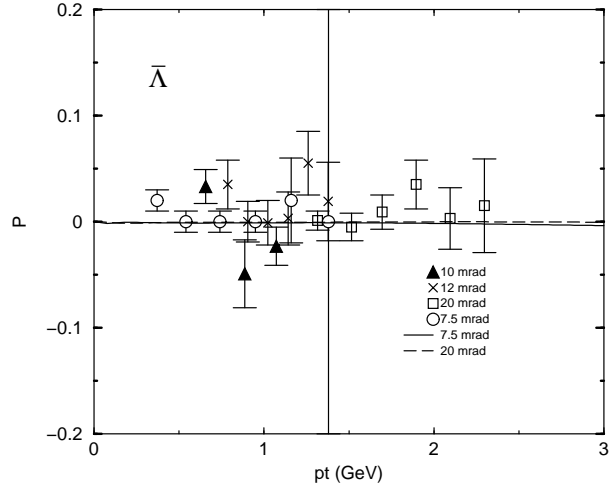}}
\caption{  ${\overline{\Lambda}}$ polarization 
 compared with the experimental data of
\cite{lam} and \cite{lu} (proton beam).}
\label{plbar}
\end{figure}

\begin{figure}[hbtp]
\centerline{
\epsfxsize=8 cm
\epsffile{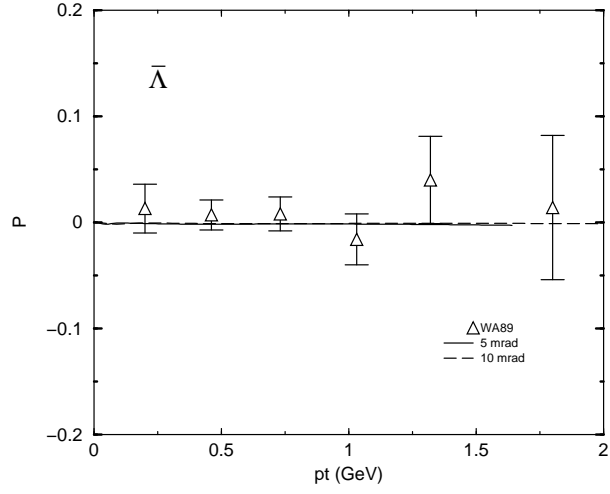}}
\caption{${\overline{\Lambda}}$ Polarization  for $\beta=2$,
 compared with the  WA89 experimental data \cite{adam}, obtained with a 
330 GeV $\Sigma^-$  beam and at $< x_F >$=0.11.}
\label{pslbar}
\end{figure}

\begin{figure}[hbtp]
\centerline{
\epsfxsize=8.cm
\epsffile{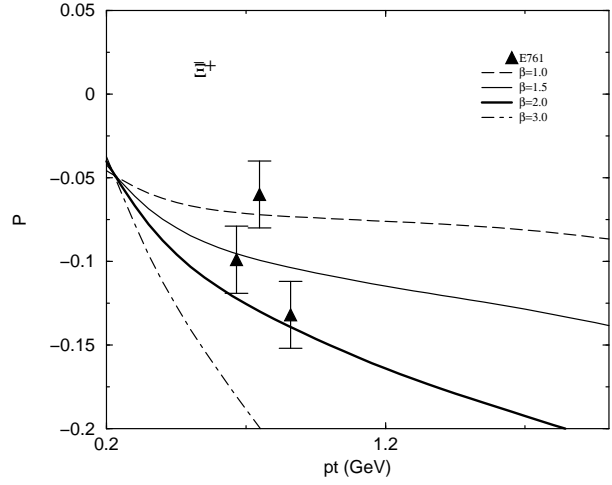}}
\caption{${\overline{\Xi}}^+$ polarization, for some values of 
$\beta$ 
compared with the E761 data  \cite{ho} (production angle of
2.4 mrad and incoming beam of 800 GeV).}
\label{px}
\end{figure}

\begin{figure}[hbtp]
\centerline{
\epsfxsize=8.cm
\epsffile{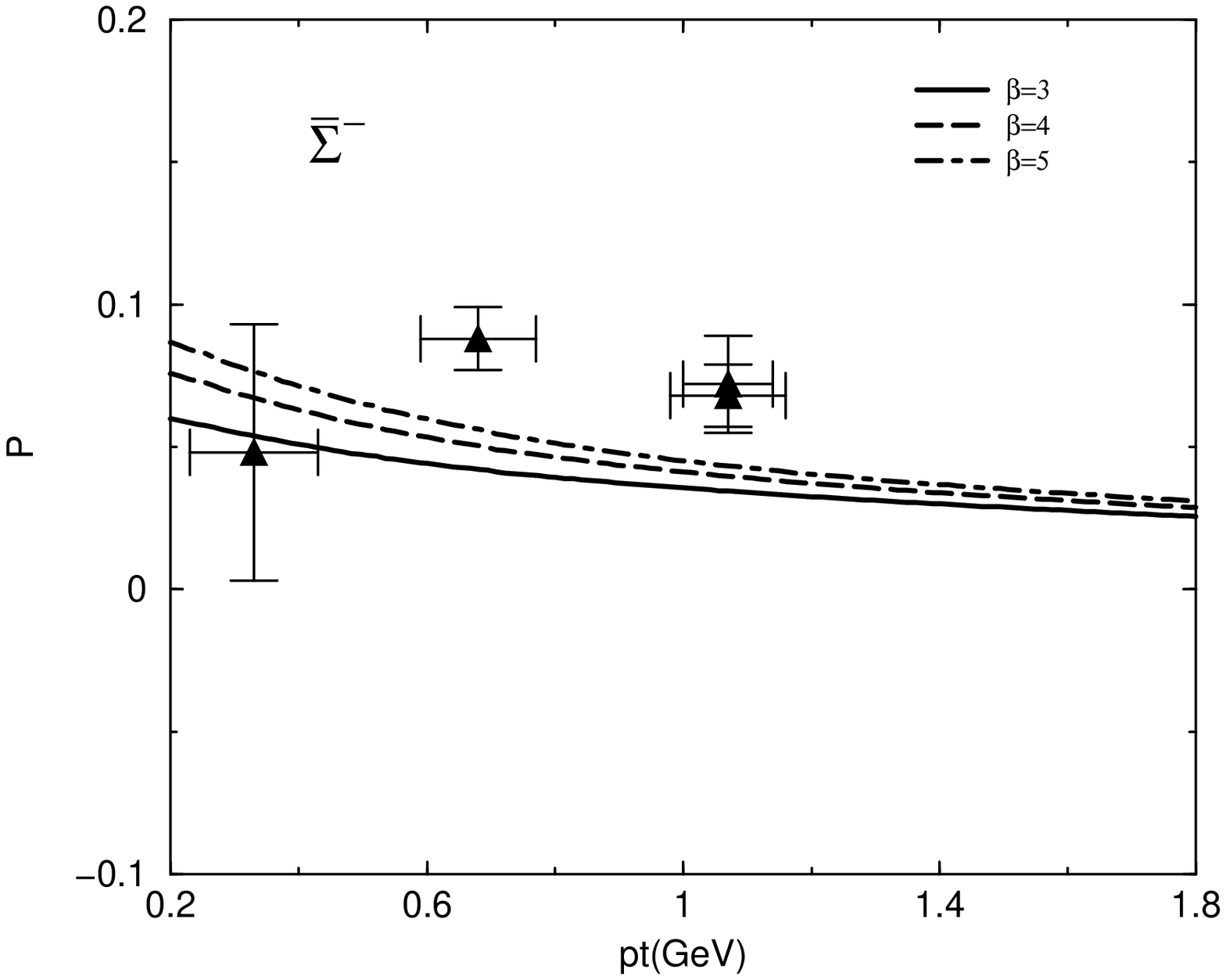}}
\caption{${\overline{\Sigma}}^-$  polarization
compared with the data of  \cite{mor} (for  $<x_F>$=0.5).}
\label{psbar}
\end{figure}

\begin{figure}[hbtp]
\centerline{
\epsfxsize=8.cm
\epsffile{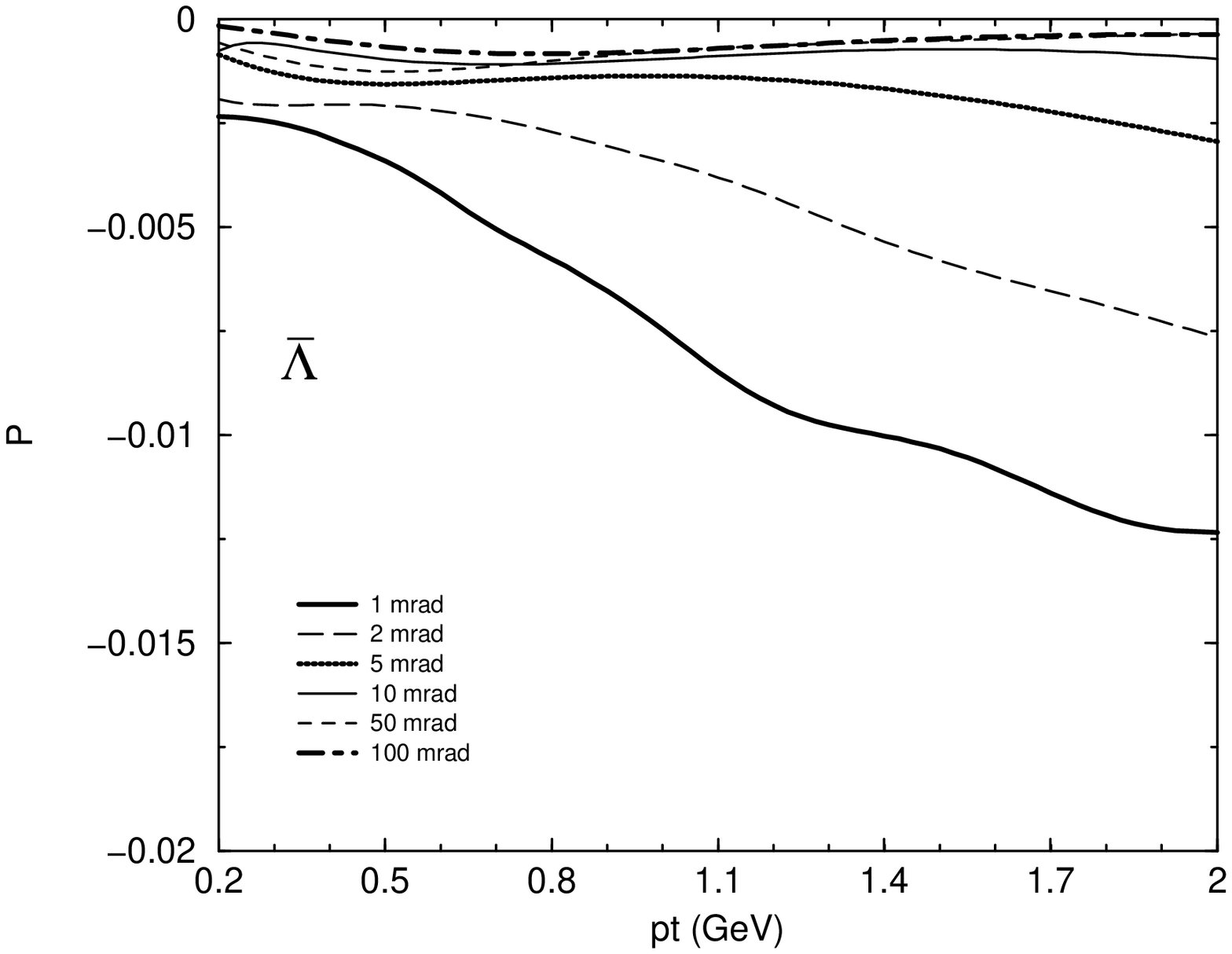}}
\caption{${\overline{\Lambda}}$ polarization, for $\beta=2$ and some production
angles}
\label{pltt}
\end{figure}

\begin{figure}[hbtp]
\centerline{
\epsfxsize=8.cm
\epsffile{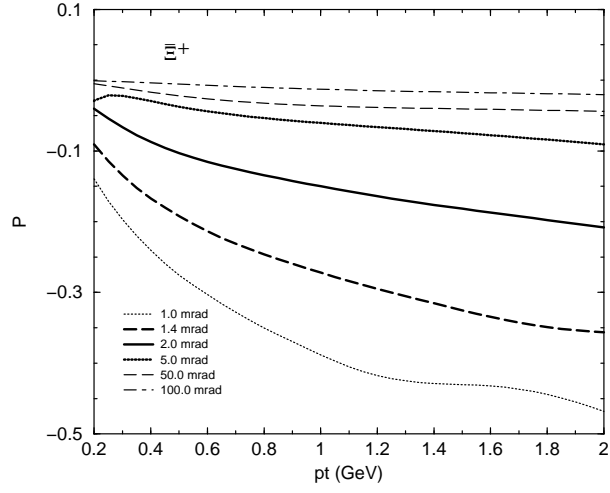}}
\caption{${\overline{\Xi}}^+$ polarization, for $\beta=2$ and some production
angles}
\label{pxtt}
\end{figure}

\begin{figure}[hbtp]
\centerline{
\epsfxsize=8.cm
\epsffile{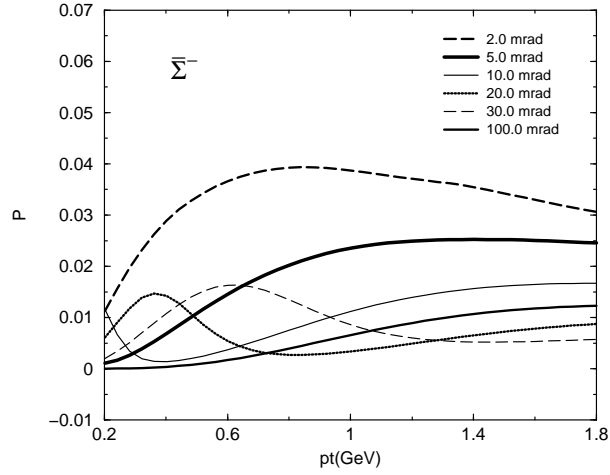}}
\caption{${\overline{\Sigma}}^-$  polarization for $\beta=3$ and some
 production angles}
\label{pstt}
\end{figure}

Integrating eq. (\ref{pm}) with the Monte Carlo method, 
we obtained the ${\overline{\Lambda}}$, ${\overline{\Sigma}^-}$ and 
${\overline{\Xi}^+}$ polarizations that are shown in Fig. 
\ref{plbar}-\ref{pstt}.

Fig. \ref{plbar} shows the ${\overline{\Lambda}}$ as a function of $p_t$,
produced from a 400 GeV proton beam and
for the production angles $\theta_L$=7.5 and 20 mrad
compared with the data from \cite{lam}, \cite{lu}
We can see that the model gives $P_{\overline{\Lambda}}\sim$0 
in the considered $p_t$ region,
what is 
consistent with the experimental results.
This fact may be justified observing Fig. \ref{pilmpol}, in an average 
calculation
the small contributions of the microscopic processes are almost totally 
canceled.

In Fig. \ref{pslbar} the calculated polarization is now compared with the data
obtained with a 330 GeV $\Sigma^-$ beam \cite{adam}, and the prediction  
$P_{\overline{\Lambda}}\sim$0 is still consistent with the data.

The calculated ${\overline{\Xi}}^+$ polarization can be seen in Fig. \ref{px}.
The resultant polarization is negative and with $\beta\sim$2, the results are 
in accord with the experimental data from the E761 experiment \cite{mor},
where $\theta_L$=2.4 mrad and $p_L$=800 GeV.

The resultant 
${\overline{\Sigma}}^-$  polarization is positive, but with magnitude smaller 
then the one of the experimental data \cite{mor} (Fig. \ref{psbar}). 
Considering that the experiment  is for $p$-Cu collisions 
one must expect a value of $\beta$ between 2.0 and 4.0 (see Fig. 
(\ref{dnpa})). 
The production angle $\theta_L$ is between
 0.9 mrad and  2.9 mrad, and we made the calculations  in this 
region, considering the average value of $x_F$ in the experiment, that is 0.5
\cite{mor}. 
In this case, we expect that some effects must be included in order to 
improve the accord with the experimental data.

Fig. \ref{pltt}-\ref{pstt}  shows the  
${\overline{\Lambda}}$, ${\overline{\Sigma}^-}$ and 
${\overline{\Xi}^+}$ polarization dependence with the production 
angle $\theta_L$. Observing the figures, we can conclude that the antihyperon 
polarization is a small angle effect, increasing $\theta_L$ the polarization 
vanishes. The model predicts $P_Y\sim$0 for $\theta_L >$50 mrad. This fact
explains why the antihyperon polarization was found experimentally in angles
as small as few mrad.

\section{Discussion of the Results}

In this paper, we proposed a model in order to explain the antihyperon 
polarization data. It is an indirect mechanism, the antihyperon is produced 
unpolarized, and then becomes polarized by the final-state interactions. 
This model supplements the direct mechanisms, that describe the hyperon
polarization, and are based on the leading particle effect 
\cite{lund}-\cite{trosh}. 
We expect that
the proposed mechanism also produces hyperons, but when compared with the 
direct mechanisms, it will be just a correction. 

The final ${\overline{\Lambda}}$ and ${\overline{\Xi}^+}$ polarizations 
agree quite well with the experimental data. 
The
${\overline{\Sigma}^-}$
polarization is  also in accord with the 
data, but in order to improve the agreement,
 some effects that we neglected during the
calculation may be taken into account. The 
$\pi{\overline{\Lambda}}\rightarrow \pi {\overline{\Sigma}}$, for example were
not included, and they may contribute to include the final 
${\overline{\Sigma}^-}$ polarization. This kind of reaction will be included 
in future works. The $\rho YY'$ coupling is another object that must be 
considered, the current value for these couplings is far from being 
well established yet. 

As we can see, the model shows that the ${\overline Y}$ polarization is a 
small 
angle effect, our prediction is that $P_{\overline{Y}}$ is very small for 
$\theta_L > 50$ mrad. Another result is that  $P_{\overline{Y}}$ does not 
depends much on the incident particle ($p$, $\Sigma$ ...). It may only change 
when the parametrization of $d\sigma / d\alpha$ changes, what would occur
specially  in $AA$ collisions.

\vspace{.6cm}

\ni {\bf Acknowledgments}

\bigskip

We would like to thank M.R. Robilotta for the discussions about the hadron 
interactions. This work was 
supported by FAPESP. 

\vfill\eject


\bigskip

\appendix

\renewcommand{\theequation}{\Alph{section}\arabic{equation}}

\setcounter{equation}{0}

\section{Basic Formalism}

In this paper $p$ and $p^{\prime}$ are the initial and final hyperon
4-momenta, $k$ and $k^{\prime}$ are the initial and final pion 4-momenta, so
the Mandelstam variables are 
\begin{eqnarray}
s &=& (p+k)^2=(p^{\prime}+k^{\prime})^2 \\
t &=& (p-p^{\prime})^2=(k-k^{\prime})^2 \\
u &=& (p^{\prime}-k)^2=(p-k^{\prime})^2 \ \ .
\end{eqnarray}

With these variables, we can define 
\begin{eqnarray}
\nu &=& {\frac{s-u}{4m}} \\
\nu_0 &=& {\frac{2m_\pi^2-t}{4m}} \\
\nu_r &=& {\frac{m_r^2-m^2-k.k^{\prime}}{2m}} \ \ ,
\end{eqnarray}

where $m$, $m_r$ and $m_\pi$ are, respectively, the hyperon mass, the 
resonance mass and the pion mass. The scattering amplitude for an isospin 
$I$ state is 
\beq
T_I=\overline{u}(\vec p')\lc A^I + 
{\frac{(\not\!k + \not\!k')}{2}}B^I \rc u(\vec p)\ ,  
\eeq

\ni where $A_I$ and $B_I$ are calculated using the Feynman diagrams. So the 
scattering matrix is 
\beq
M_I^{ba} = {\frac{T_I^{ba}}{8\pi\sqrt{s}}} = f_I(\theta) + \vec\sigma.\hat n
g_I(\theta) = f_1^I + {\frac{(\vec\sigma .\vec k' )(\vec\sigma .\vec k)}{kk'}}f_2^I \ \ , 
\eeq

with 
\begin{eqnarray}
& &f_1^I(\theta) = {\frac{(E+m)}{8\pi\sqrt{s}}} \lb A_I + (\sqrt{s}%
-m)B_I\rb\ \ , \\
& & f_2^I(\theta) = {\frac{(E-m)}{8\pi\sqrt{s}}} \lb -A_I + (\sqrt{s}%
+m)B_I\rb \ \ ,
\end{eqnarray}

where $E$ is the hyperon energy. The partial-wave decomposition is done
with 
\beq
a_{l\pm} = {\frac{1}{2}}\int_{-1}^{1}\lb P_l(x)f_1(x) + P_{l\pm
1}(x)f_2(x)  \rb dx \ \ . 
\eeq

In our calculation (tree level) $a_{l\pm}$ is real. With the unitarization, 
as explained in Section III, we obtain 
\beq
 a_{l\pm}^U = {\frac{1}{2ik}}\lb e^{2i\delta_{l\pm}}-1\rb 
 = \frac{e^{i\delta_{l\pm}}}{k}\,{\rm sen}(\delta_{l\pm})
 \rightarrow a_{l\pm} \ \ . 
\eeq

These complex amplitudes are used to calculate
\begin{eqnarray}
&& f(\theta)=\sum_{l=0}^{\infty}\lbrack (l+1)a_{l+}+la_{l-}\rbrack\,P_l(x)
 \ , \\
&& g(\theta)=i\sum_{l=1}^{\infty}\lbrack a_{l+}-a_{l-}\rbrack\,P_l^{(1)}(x)
 \ . 
\end{eqnarray}

We have, then, in the center-of-mass frame, 
\begin{eqnarray}
& {\frac{d\sigma}{d\Omega}} & = |f|^2 + |g|^2 \ , \\
& {d\sigma\over dt} & = {\pi\over k^2} \lb |f|^2 + |g|^2  \rb \ , \\
& \vec P & =-2{\frac{{\rm Im}(f^*g)}{|f|^2 + |g|^2}}\hat n \ , \\
& \sigma_t & = 4\pi \sum_l\lb (l+1)|a_{l+}|^2 + l|a_{l-}|^2\rb \ .
\end{eqnarray}

The $Y\pi Y^\prime$ coupling constants may be obtained considering the  SU(3) 
relations 
\cite{pilk}
\bea
g_{\pi NN}&=& g \nn  \\
g_{\Lambda\pi\Sigma}&=& {2\over \sqrt{3}}g\ \alpha \nn  \\
g_{\Sigma\pi\Sigma}&=& 2g(1- \alpha) \nn  \\
g_{\Xi\pi\Xi}&=& g(1-2\alpha) \  \  ,
\eea

\ni
considering the current value $g_{\pi NN}$=13.40 and 
$g_{\Lambda\pi\Sigma}$=12.92, obtained in \cite{lois} from the study of the 
hypernuclei, we have $\alpha$=0.83, and the other coupling constants are 
determined. Similar relations hold for the $Y\rho Y^\prime$ couplings. The 
$Y\pi Y^*$ couplings are calculated with the procedure described in \cite{cy}.

\vfill\eject


\end{document}